\PassOptionsToPackage{numbers}{natbib}
\PassOptionsToPackage{bookmarks=true,breaklinks=true,letterpaper=true,colorlinks,citecolor=blue,linkcolor=blue,urlcolor=blue}{hyperref}
\documentclass[nonacm,sigplan]{acmart}

\usepackage{listings}
\usepackage{xcolor}
\usepackage{multicol}
\usepackage{multirow}
\usepackage[inkscapelatex=false]{svg}
\usepackage{graphicx} 
\usepackage{parcolumns}
\usepackage{enumitem}
\usepackage{xurl}
\usepackage{tikz}
\newcommand\encircle[1]{%
  \tikz[baseline=(X.base)] 
    \node (X) [draw, shape=circle, inner sep=0, fill=black, text=white] {\strut #1};%
}

\usepackage{float}
\usepackage{tabularx}
\usepackage{caption}
\lstdefinestyle{mystyle}{
    backgroundcolor=\color{white},   
    commentstyle=\color{green},
    keywordstyle=\color{blue},
    numberstyle=\tiny\color{gray},
    stringstyle=\color{orange},
    basicstyle=\linespread{0.6}\ttfamily\footnotesize,
    breakatwhitespace=false,             
    breaklines=true,                     
    captionpos=b,                        
    keepspaces=true,                     
    numbers=left,                        
    numbersep=5pt,                       
    showspaces=false,                    
    showstringspaces=false,              
    showtabs=false,                      
    tabsize=2,
    aboveskip=2pt,
    belowskip=2pt
}

\newcolumntype{L}{>{\raggedright\arraybackslash}X} 
\newcolumntype{C}[1]{>{\centering\arraybackslash}m{#1}} 

\newcolumntype{Y}{>{\centering\arraybackslash}X}
\newcolumntype{W}{>{\hsize=1.4\hsize\linewidth=\hsize\centering\arraybackslash}X}
\newcolumntype{N}{>{\hsize=0.03\hsize\linewidth=\hsize\centering\arraybackslash}X
}
\newcolumntype{M}{>{\hsize=0.4\hsize\linewidth=\hsize\centering\arraybackslash}X}

\newcolumntype{A}{>{\hsize=1.2\hsize\linewidth=\hsize\centering\arraybackslash}X}
\newcolumntype{B}{>{\hsize=0.87\hsize\linewidth=\hsize\centering\arraybackslash}X
}
\newcolumntype{C}{>{\hsize=1.1\hsize\linewidth=\hsize\centering\arraybackslash}X}

\newcolumntype{P}[1]{>{\centering\arraybackslash}p{#1}}

\title{DP-HLS: A High-Level Synthesis Framework for Accelerating Dynamic Programming Algorithms in Bioinformatics}

\author{Yingqi Cao}
\authornote{Both authors contributed equally to this research.}
\affiliation{%
  \institution{University of California San Diego}
  \city{San Diego}
  \state{CA}
  \country{USA}
}

\author{Anshu Gupta}
\authornotemark[1]
\affiliation{%
  \institution{University of California San Diego}
  \city{San Diego}
  \state{CA}
  \country{USA}
}

\author{Jason Liang}
\affiliation{%
  \institution{University of California San Diego}
  \city{San Diego}
  \state{CA}
  \country{USA}
}

\author{Yatish Turakhia}
\affiliation{%
  \institution{University of California San Diego}
  \city{San Diego}
  \state{CA}
  \country{USA}
}

\begin{document}

\begin{abstract}
\sloppy

Dynamic programming (DP) based algorithms are essential yet compute-intensive parts of numerous bioinformatics pipelines, which typically involve populating a 2-D scoring matrix based on a recursive formula, optionally followed by a traceback step to get the optimal alignment path. DP algorithms are used in a wide spectrum of bioinformatics tasks, including read assembly, homology search, gene annotation, basecalling, and phylogenetic inference. So far, specialized hardware like ASICs and FPGAs have provided massive speedup for these algorithms. However, these solutions usually represent a single design point in the DP algorithmic space and typically require months of manual effort to implement using low-level hardware description languages (HDLs). This paper introduces DP-HLS, a novel framework based on High-Level Synthesis (HLS) that simplifies and accelerates the development of a broad set of bioinformatically relevant DP algorithms in hardware. DP-HLS features an easy-to-use template with integrated HLS directives, enabling efficient hardware solutions without requiring hardware design knowledge. In our experience, DP-HLS significantly reduced the development time of new kernels (months to days) and produced designs with comparable resource utilization to open-source hand-coded HDL-based implementations and performance within 7.7–16.8\% margin. DP-HLS is compatible with AWS\textsuperscript{\textregistered} EC2 F1 FPGA instances. To demonstrate the versatility of the DP-HLS framework, we implemented 15 diverse DP kernels, achieving 1.3–32$\times$ improved throughput over state-of-the-art GPU and CPU baselines and providing the first open-source FPGA implementation for several of them. The DP-HLS codebase is available freely under the MIT license at \texttt{\href{https://github.com/TurakhiaLab/DP-HLS}{https://github.com/TurakhiaLab/DP-HLS}} and its detailed wiki at \texttt{\href{https://turakhia.ucsd.edu/DP-HLS/}{https://turakhia.ucsd.edu/DP-HLS/}}. 
\end{abstract}
\maketitle 
\pagestyle{plain} 


\section{Introduction}
\sloppy

Genomic data is one of the fastest-growing data types globally, far outpacing Moore’s law in terms of data generation \cite{stephens2015big}. To meet the rising computational demands of analyzing and interpreting this data, several efforts have focused on accelerating bioinformatics applications on hardwares like GPUs, FPGAs, and ASICs  \cite{lindegger2023scrooge, ahmed2020gpu, zeni2020logan, ahmed2019gasal2, goenka2020segalign, park2024agatha,aguado2022wfa, zenigpu2024, fei2018fpgasw, turakhia2018darwin, turakhia2019darwin, haghi2021fpga, zhang2007implementation, msa_dp, bsw_systolic, seedex, dp_2_piece, dp_protein_affine, bwamem_accel, msa_fpga, huang2017hardware, walia2024talco, cali_genasm_2020, cali2022segram}.

While a lot of these accelerators are custom solutions to target a narrow application in bioinformatics, they also share notable similarities. For example, many of these solutions accelerate an algorithm based on dynamic programming (DP) \cite{bellman1966dynamic}. This is unsurprising, as DP provides an efficient framework for comparing biological sequences—such as DNA, RNA, proteins, or even electrical signals from sequencing instruments—which is fundamental to many bioinformatics tasks, such as local pairwise alignments \cite{smith1981identification, ukkonen1985finding}, multiple sequence alignment \cite{clustalomega, edgar2004muscle}, homology searches \cite{eddy1998profile, altschul1990basic}, whole-genome alignments \cite{harris2007improved}, basecalling \cite{simpson2017detecting}, and variant calling \cite{nielsen2012snp}. DP algorithms are computationally intensive, and therefore, they often dominate the runtime of these applications \cite{subramaniyan2021genomicsbench}. Recognizing their importance to bioinformatics, NVIDIA\textsuperscript{\textregistered} recently introduced a specialized instruction, DPX, specifically to accelerate DP algorithms on GPUs~\cite{elster2022nvidia}.

Another key characteristic, particularly in FPGA and ASIC solutions \cite{fei2018fpgasw, turakhia2018darwin, turakhia2019darwin, haghi2021fpga, zhang2007implementation, msa_dp, bsw_systolic, seedex, dp_2_piece, dp_protein_affine, bwamem_accel, msa_fpga, huang2017hardware, walia2024talco, cali_genasm_2020, cali2022segram}, is the use of a hardware primitive—linear systolic arrays—that has been recognized since the 1980s for its efficiency in accelerating DP algorithms \cite{lipton1985systolic}. Most of these solutions focus on a specific or narrow set of 2-D DP algorithms and are typically designed at the Register Transfer Level (RTL) using low-level Hardware Description Languages (HDLs) like Verilog or VHDL, which makes them difficult to design and modify. A similar observation was made by a recent work, GenDP \cite{gu2023gendp}, which proposed a linear systolic array with software-programmable processing elements to accelerate a broad range of DP algorithms. However, software-programmable solutions introduce significant overhead on circuit-programmable FPGAs, which are the primary target of our work since they have already found commercial applications in bioinformatics \cite{behera2024comprehensive, goyal2017ultra, timelogic}.


\begin{figure*}
    \includegraphics[width=0.80\textwidth]{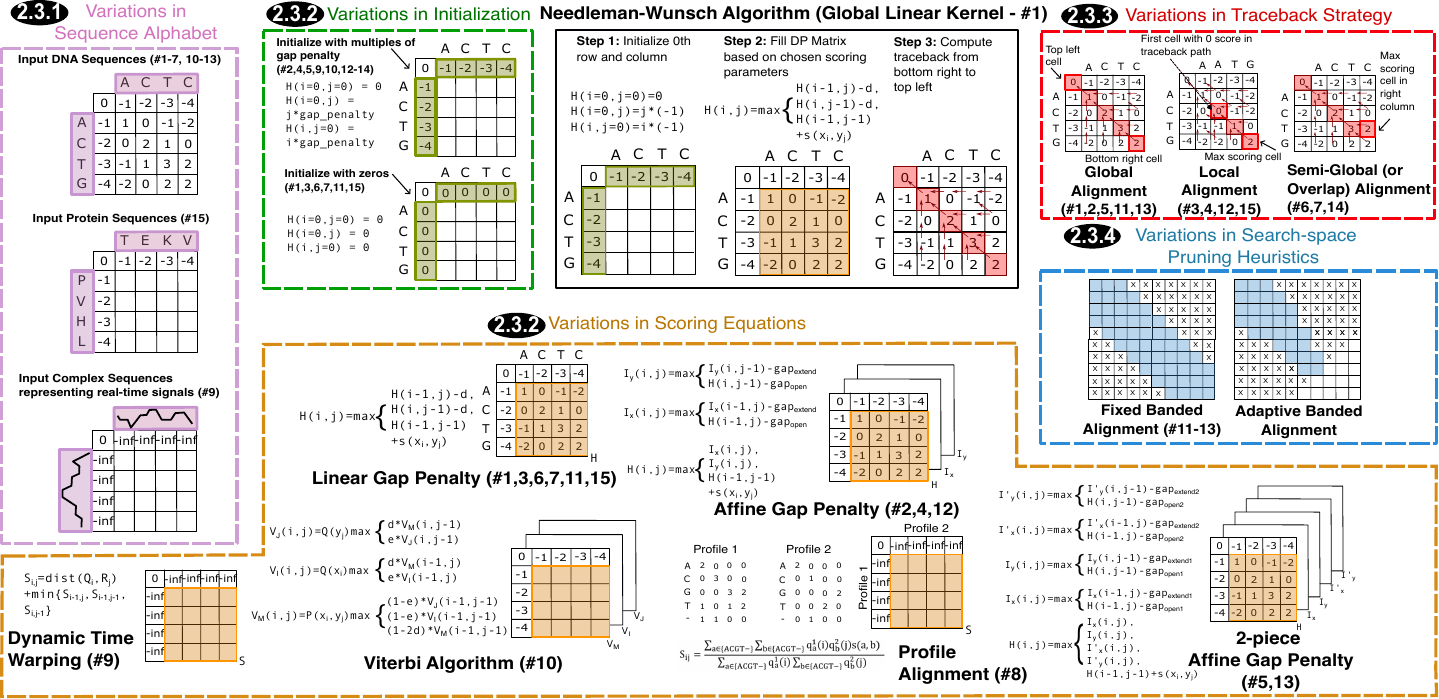}
    \caption{Common variations in 2-D DP algorithms in Bioinformatics. Kernels are indexed using \#'s based on Table~\ref{table:summary}.}
    \label{fig:background}
\end{figure*}

In this paper, we present \emph{DP-HLS}, a novel framework based on High-Level Synthesis (HLS) for accelerating broad DP kernels on FPGAs. A key feature of the framework is the separation of \emph{back-end} optimizations from the \emph{front-end} interface, enabling users to define new kernels in C++ without needing expertise in HLS or digital design. DP-HLS makes the following key contributions:

\begin{enumerate}[leftmargin=*]
\item We developed DP-HLS, an HLS framework that, by separating front-end and back-end, introduces a layer of abstraction in the HLS flow to significantly enhance the productivity in deploying DP kernels on FPGAs. Specifically, the front-end provides a high degree of flexibility to specify new DP kernels in C\texttt{++} without needing to add or modify HLS directives, while the back-end uses fixed HLS directives to efficiently translate these specifications into optimized RTL implementations with systolic arrays. Our experiments confirm that the generated RTL code exhibits the expected linear systolic array behavior.
\item Using the DP-HLS framework, we implemented 15 bioinformatically relevant DP kernels on FPGAs, covering a wide range of applications from basecalling to protein sequence alignment. All kernels are functionally verified and deployed on AWS\textsuperscript{\textregistered} EC2 F1 instances for broad accessibility. For most kernels, no open-source FPGA implementations are available.
\item For the three DP kernels with available hand-optimized RTL implementations, DP-HLS achieved throughput within 7.7-16.8\% and comparable resource utilization. However, DP-HLS significantly boosts design productivity, from months to days, compared to manual RTL design.
\item For several DP kernels implemented in state-of-the-art CPU and GPU libraries, DP-HLS delivered a 1.3–32$\times$ improvement in throughput over CPU- and GPU-optimized AWS\textsuperscript{\textregistered} EC2 instances of the same cost.
\item We demonstrate, through an example, that recently proposed tiling heuristics \cite{turakhia2018darwin} are compatible with DP-HLS and can be used for performing both short and long sequence alignments on the FPGA.
\item We have made the entire framework publicly available, including
all 15 DP kernels as case studies at \texttt{\href{https://github.com/TurakhiaLab/DP-HLS}{https://github.com/TurakhiaLab/DP-HLS}}. Additionally, we provide a comprehensive wiki to assist new users at \texttt{\href{https://turakhia.ucsd.edu/DP-HLS/}{https://turakhia.ucsd.edu/DP-HLS/}}.

\end{enumerate}



\section{Background and Motivation}
\label{sec:background}

\begin{table*}[h!]
\tiny
\centering
\begin{tabularx}{\textwidth}{|N|M|W|A|B|C|}
  \hline
  \textbf{\#} & \textbf{Alphabet} & \textbf{2-D DP Kernel} & \textbf{State-of-the-art Tools} & \textbf{Example Applications} & \textbf{Modifications in DP-HLS} \\ 
  \hline
  1 & DNA & Global Linear Alignment (Needleman-Wunsch)\label{kernel:1}  & BLAST\cite{ncbi_blast}, EMBOSS Stretcher\cite{PMID:38597606} & Similarity Search & N/A\\
  \hline
  2 & DNA & Global Affine Alignment (Gotoh)\label{kernel:2} & BLAST\cite{ncbi_blast}, EMBOSS Needle\cite{PMID:38597606} & Accurate Similarity Search & Scoring \\
  \hline
  3 & DNA & Local Linear Alignment (Smith-Waterman)\label{kernel:3} & BLAST\cite{ncbi_blast}, FASTA\cite{fasta}, BLAT\cite{kent2002blat} & Homology Search & Initialization and Traceback \\
  \hline
  4 & DNA & Local Affine Alignment (Smith-Waterman-Gotoh)\label{kernel:4} & BLAST\cite{ncbi_blast}, LASTZ\cite{lastz} & Whole Genome Aligment & Scoring, Initialization and Traceback\\
  \hline
  5 & DNA & Global Two-piece Affine Alignment\label{kernel:5} & Minimap2\cite{li_minimap2_2018} & Long Read Alignment & Scoring\\
  \hline
  6 & DNA & Overlap Alignment\label{kernel:6} & CANU\cite{canu}, Flye\cite{flye} & Genome Assembly & Initialization and Traceback \\
  \hline
  7 & DNA & Semi-global Alignment\label{kernel:7} & BWA-MEM\cite{bwa-mem} & Short Read Alignment & Initialization and Traceback\\
  \hline
  8 & Seq. Profiles & Profile Alignment\label{kernel:8} & CLUSTALW\cite{thompsonCLUSTALImprovingSensitivity1994}, MUSCLE\cite{edgar2004muscle} & Multiple Sequence Alignment & Sequence Alphabet and Scoring \\
  \hline
  9 & Complex Nos. & Dynamic Time Wrapping Algorithm (DTW)\label{kernel:9} & SquiggleKit\cite{squigglekit} & Basecalling & Sequence Alphabet and Scoring \\
  \hline
  10 & DNA & Viterbi Algorithm (PairHMM)\label{kernel:10} & HMMER\cite{finnHMMERWebServer2011}, AUGUSTUS\cite{stanke2005augustus} & Remote Homology Search, Gene Prediction & Scoring (no Traceback) \\
  \hline
  11 & DNA & Banded Global Linear Alignment\label{kernel:11} & BLAST\cite{ncbi_blast}, Bowtie\cite{bowtie} & Fast Similarity Search & Scoring and Initialization \\
  \hline
  12 & DNA & Banded Local Affine Alignment\label{kernel:12} &  Minimap2\cite{li_minimap2_2018} & Long Read Assembly & Initialization, Scoring (no Traceback) \\
  \hline
  13 & DNA & Banded Global Two-piece Affine Alignment\label{kernel:13} & Minimap2\cite{li_minimap2_2018} & Long Read Assembly & Scoring, Initialization and Traceback\\
  \hline
  14 & Integers & Semi-global DTW (sDTW)\label{kernel:14} & SquiggleFilter\cite{dunn2021squigglefilter}, RawHash\cite{rawhash} & Basecalling & Sequence Alphabet and Scoring \\
  \hline
  15 & Amino acids & Local Linear Alignment with protein sequences\label{kernel:15} & EMBOSS Water\cite{PMID:38597606}, BLASTp\cite{ncbi_blast}, DIAMOND\cite{buchfink2021sensitive} & Protein Sequence Alignment & Sequence Alphabet and Scoring \\
  \hline
\end{tabularx}
\caption{Common bioinformatics kernels using 2-D DP algorithms, along with their associated tools, applications, and modifications relative to the baseline kernel (\hyperref[kernel:1]{\#1}: Global Linear Alignment). Some tools implement multiple kernels. } 
\label{table:summary}
\end{table*}


\subsection{The 2-D Dynamic Programming Paradigm}
Biological sequences, including DNA, RNA, and proteins, are the building blocks of life. DNA and RNA consist of four nucleotides: \texttt{A}, \texttt{C}, \texttt{G}, and \texttt{T} (\texttt{U} in RNA), while proteins are made up of 20 amino acids. Many bioinformatics problems involve comparing these biological sequences (DNA, RNA, proteins) to identify similarities and differences \cite{pal2006evolutionary, cohen2004bioinformatics}. A common approach is using 2-D dynamic programming (DP) algorithms (2-D DP paradigm), which typically include: i) \emph{initialization}, ii) \emph{matrix fill}, and iii) optional \emph{traceback} \cite{needleman1970general,smith1981identification}. The \emph{initialization} step arranges the two sequences on a 2-D grid, called \emph{DP matrix}, with the first row and column having predefined scores.  The \emph{matrix fill} step uses a recursive formula to score each cell based on its three neighboring cells: above, left, and diagonal (Fig. \ref{fig:background}). Lastly, the \emph{traceback} step recovers the path in the DP matrix corresponding to the optimal score. Without traceback, only the optimal score is returned. Fig. \ref{fig:background} illustrates these steps using the most simplest 2-D DP algorithm, the Needleman-Wunsch algorithm \cite{needleman1970general}.


\subsection{Variations in the 2-D DP Paradigm}

Table \ref{table:summary} shows a selection of 2-D DP kernels (indexed with ‘\#’ hereafter) that are commonly used in various bioinformatics applications, widely cited, and in some cases, targeted by hardware accelerators.
We summarize key kernel characteristics and direct readers to original sources for details. These variations fall into four main categories (Fig. \ref{fig:background}): i) \textit{Sequence Alphabet}, ii) \textit{Scoring Equations}, iii) \textit{Traceback Strategy}, and iv) \textit{Search-space Pruning Heuristics}, as outlined below.

\subsubsection{Sequence Alphabet:} \label{sec:alphabet} An alphabet refers to the set of characters used to represent sequences, such as DNA, RNA, or protein, which typically consist of 4 or 20 characters, though some variations that introduce ambiguous bases, e.g., \texttt{N}s, also exist \cite{ncbi_blast, harris2007improved}. In Profile Alignment (\hyperref[kernel:8]{\#8}), multiple DNA (protein) sequences are aligned with each other where each character is represented as a tuple of 5 (21) integers, referring to the frequencies of the 4 (20) nucleotides (amino acids) and gaps at each position of the alignment \cite{wangScoringProfiletoprofileSequence2004} (Fig. \ref{fig:background}). Dynamic Time Warping (DTW) (\hyperref[kernel:9]{\#9} and \hyperref[kernel:14]{\#14}), developed to compare two time-series signals \cite{bellman1959adaptive, han2018accurate}, uses real or complex numbers (Fig. \ref{fig:background}) when used for basecalling in nanopore sequencing applications \cite{dunn2021squigglefilter}. 

\subsubsection{Scoring Equations:} \label{sec:scoring} In the 2-D DP paradigm, scoring refers to the recurrence equations that calculate cell scores,  by typically rewarding matches, and penalizing mismatches or gaps. We mention some of the scoring strategies and their common variations below.


    \indent \textbf{(a) Substitution Scores:} The simplest form of scoring uses a single value to reward a match or penalize a mismatch, or both. However, due to varying substitution frequencies (e.g., transitions vs. transversions), larger substitution matrices of penalty values are also common for DNA, RNA, or protein alignments \cite{henikoff_amino_1992} (Fig. \ref{fig:background}). In Profile Alignments (\hyperref[kernel:8]{\#8}) and DTW (\hyperref[kernel:9]{\#9}), substitution values are computed dynamically using metrics like Sum-of-Pairs scoring \cite{wangScoringProfiletoprofileSequence2004} and Manhattan/Euclidean distance \cite{dunn2021squigglefilter,skutkova_classification_2013}, respectively.
    
    \textbf{(b) Gap Penalties:} In sequence alignment, gap penalties penalize gaps (insertions or deletions). A linear gap penalty applies a constant penalty for every gap, while an affine gap penalty better models biological reality by assigning  a higher penalty to opening a gap than extending an existing gap \cite{gotoh_improved_1982}. This requires three score values (\texttt{H, I, D}) per DP-matrix's cell. Minimap2 \cite{li_minimap2_2018} uses a two-piece affine gap penalty with five values per cell to further improve the distinction between biological gaps and sequencing errors. In general, each affine layer adds two values to the DP matrix for gap cost to resemble a smooth convex function. 
    
    
    \textbf{(c) Initialization:} The scoring equations also define how the initial row and column are scored. Depending on which traceback strategy is used, the scores could be a constant (e.g., 0 or -$\infty$) or a function of the gap penalties (Fig. \ref{fig:background}). 
    
    \textbf{(d) Min/Max Objective Function:} Most DP formulations (\hyperref[kernel:1]{\#1}-\hyperref[kernel:7]{7}, \hyperref[kernel:11]{\#11}-\hyperref[kernel:13]{13}) penalize the gap and find the maximum cell score, whereas DTW aims to find the minimum cell score while the score represents distances between the sequence. In this case, we need to replace the \texttt{max} function in recurrence equations with \texttt{min} (Fig. \ref{fig:background}). 

\subsubsection{Traceback Strategy: } \label{sec:traceback_strategy} 
The traceback step finds the actual alignment with the optimal score. While recurrence equations specify optimal transitions on a traceback path, the traceback strategy determines \emph{where to start and end the path}. Four strategies are common: global, local, semi-global, and overlap, each influencing the recurrence and initialization equations (Fig. \ref{fig:background}). \emph{Global} strategy performs end-to-end whole genome sequence comparison, with a traceback path from the bottom-right to the top-left cell of the DP matrix. \emph{Local} strategy finds similar subsequences and is used to identify conserved motifs or functional regions in sequences, with the traceback from the highest-scoring cell to a 0-scoring cell. \emph{Semi-global} strategy matches one sequence end-to-end with a subsequence of the other, with the traceback starting from the bottom row's highest-scoring cell to the top row.  \emph{Overlap} strategy, used in genome assembly, matches sequence ends (prefixes or suffixes) with traceback path from the highest-scoring cell in the rightmost column (bottom row) to the top row (leftmost column) of the 2-D DP matrix.


\subsubsection{Search-space Pruning Heuristics:} As 2-D DP matrices grow quadratically with sequence lengths, many algorithms employ heuristics like fixed or adaptive banding to prune unpromising regions (Fig. \ref{fig:background}). \textit{Fixed} banding algorithms compute cells near the main diagonal within a fixed distance \cite{chaoAligningTwoSequences1992}, while \textit{adaptive} methods like X-Drop \cite{zhangGreedyAlgorithmAligning2000} adjust band size dynamically based on cell scores.


\section{An Overview of the DP-HLS Framework}
\label{sec:overview}

\begin{figure*}
    \includegraphics[width=\textwidth]{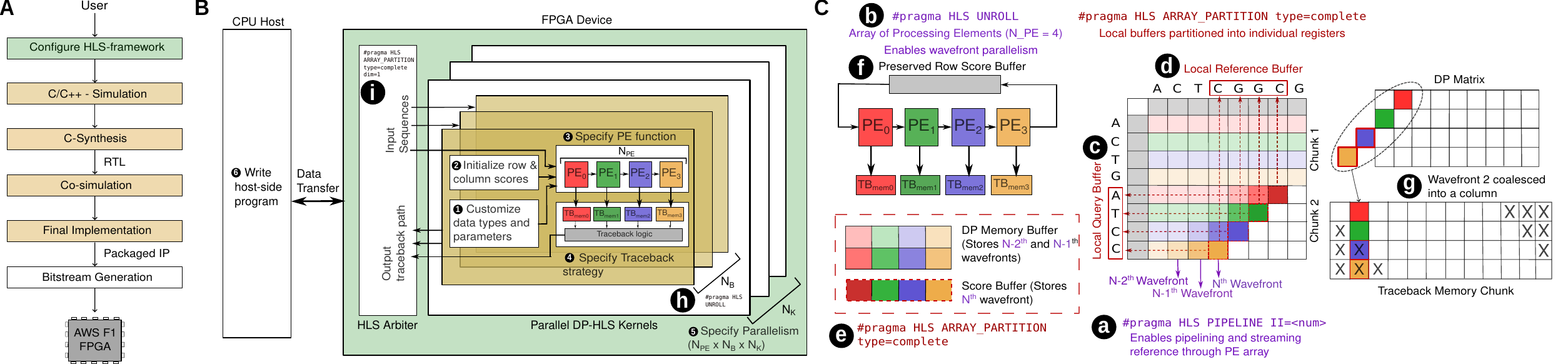}
    \caption{\textbf{DP-HLS implementation overview}. \textbf{(A)} Basic workflow of DP-HLS kernels from user customization to FPGA deployment. \textbf{(B)} Front-end layout of DP-HLS kernels with customizable modules. \textbf{(C)} Key back-end optimizations with corresponding \texttt{pragma} directives.}
    \label{fig:implementation}
\end{figure*}

DP-HLS is a framework that accelerates 2-D DP algorithms on FPGAs using HLS (Fig. \ref{fig:implementation}). Built with AMD\textsuperscript{\textregistered} Vitis HLS \cite{VitisHighLevelSynthesis2021} tool, it offers customizability for creating FPGA-accelerated kernels tailored to specific applications. The framework consists of two components: the \emph{front-end} (Section \ref{sec:front-end}), where users can specify kernels in C/C\texttt{++} without HLS expertise, supporting co-simulation, verification, and FPGA deployment; and the \emph{back-end} (Section \ref{sec:back-end}), which uses HLS directives to optimize the design for efficient hardware mapping. DP-HLS generalizes well across various 2-D DP kernels (e.g., to all kernels in Table \ref{table:summary}), including the ones dealing with long sequence alignments.
\section{Front-end Implementation}
\label{sec:front-end}


This section details the front-end of the DP-HLS framework, using examples from the 15 2-D DP kernels in Table \ref{table:summary} we implemented. Fig. \ref{fig:implementation}A shows the design flow, involving kernel configuration, simulation, synthesis, co-simulation, and final implementation. The framework also provides debugging scripts and instructions for AWS\textsuperscript{\textregistered} EC2 F1 FPGA deployment. Below, we outline  the six required steps for configuring the front-end for specific kernels.


\noindent \textbf{{\encircle{1}} Customizing Data Types and Parameters:} 
The DP-HLS framework supports custom data types of variable precision for scoring, traceback, and logic indices, enabling users to optimize efficiency for their specific kernel requirements. It also allows customization of scoring and input parameters. Primary front-end customizations are enumerated below.

    1. \textbf{Sequence Alphabets: }
   As explained in Section \ref{sec:alphabet}, the sequence alphabet in 2-D DP kernels requires high configurability. DP-HLS addresses this by allowing users to define a custom data type, \texttt{char\_t}, for the sequence alphabet. Listing \ref{listing:alphabet} (left) shows a 2-bit unsigned integer used as \texttt{char\_t} to represent the four nucleotide bases in DNA and RNA. Listing \ref{listing:alphabet} (right) shows an alphabet for the DTW kernel (\hyperref[kernel:9]{\#9}), where \texttt{char\_t\_st} is a struct of two 32-bit fixed-point numbers representing the real and imaginary parts of complex temporal signals being compared.

\begin{figure}[!htb]
\captionsetup{type=lstlisting}

\noindent\begin{minipage}{.20\textwidth}
\begin{lstlisting}[language=C++, numbers=none]
typedef ap_uint<2> char_t;
\end{lstlisting}
\end{minipage}\hfill
\begin{minipage}{.27\textwidth}
\begin{lstlisting}[language=C++,numbers=none]
struct char_t_st { ap_fixed<32, 26> real, imag; };
typedef char_t_st char_t;
\end{lstlisting}
\end{minipage}
\caption{Input alphabet for \textbf{(Left)} DNA sequences and \textbf{(Right)} complex number sequences.}
\label{listing:alphabet}
\end{figure}

    2. \textbf{Scoring Layers and Data Types:}  As described in Section \ref{sec:scoring}, some 2-D DP kernels involve multiple recurrence equations, each computing a unique value per cell. DP-HLS provides a variable, called \texttt{N\_LAYERS}, in the front-end, which configures the number of unique values computed and stored per cell of the DP matrix. So, for kernels with an affine gap penalty (Kernels \hyperref[kernel:2]{\#2}, \hyperref[kernel:4]{\#4}, \hyperref[kernel:10]{\#10}, \hyperref[kernel:12]{\#12}), \texttt{N\_LAYERS} is set to 3, for those with a two-piece affine gap penalty (Kernels \hyperref[kernel:5]{\#5}, \hyperref[kernel:13]{\#13}), it is set to 5, and for the remaining kernels, it is set to 1.

    3. \textbf{Scoring Parameters:} DP-HLS allows users to define an arbitrary number of scoring parameters with any data type in a C/C\texttt{++} struct, called \texttt{ScoringParams}, with values set at runtime by the host code. Listing \ref{listing:scoring} (left) shows an example with three parameters (\texttt{match}, \texttt{mismatch}, and \texttt{linear\_gap}) used by the Global Linear kernel (\hyperref[kernel:1]{\#1}).  The Viterbi kernel (\hyperref[kernel:10]{\#10}) uses three states (\texttt{M}, \texttt{I}, and \texttt{D}) in the Hidden Markov Model and requires a total of 27 parameters. Two parameters, $\mu$ and $\lambda$, store the transition probabilities between the three hidden states, and a 5$\times$5 \texttt{emission} matrix stores the emission probabilities for all pairs of characters (\texttt{A}, \texttt{C}, \texttt{G}, \texttt{T}, and \texttt{–}) in the \texttt{M} state (Listing \ref{listing:scoring} (right)). 
    
\begin{figure}
\captionsetup{type=lstlisting}
\noindent\begin{minipage}{.22\textwidth}
\begin{lstlisting}[language=C++, numbers=none]
struct ScoringParams {
    type_t mismatch;
    type_t match;
    type_t linear_gap;
} params;
\end{lstlisting}
\end{minipage}\hfill
\begin{minipage}{.25\textwidth}
\begin{lstlisting}[language=C++, numbers=none]
struct ScoringParams {
    type_t log_mu;
    type_t log_lambda;
    type_t emission[5][5];
} params;
\end{lstlisting}
\end{minipage}
\caption{Definition of \texttt{ScoringParams} used in \textbf{(Left)} Global Linear kernel (\hyperref[kernel:1]{\#1}) and \textbf{(Right)} Viterbi kernel (\hyperref[kernel:10]{\#10}).}
\label{listing:scoring}
\end{figure}

    4. \textbf{Maximum Sequence Lengths:} In DP-HLS, users can define maximum reference and query sequence lengths using \texttt{MAX\_REFERENCE\_LENGTH} and \texttt{MAX\_QUERY\_LENGTH}, which help determine the memory sizes for storing sequences and traceback pointers on the FPGA device. While the kernel supports fixed maximum lengths, longer sequences can be handled using software tiling approaches \cite{turakhia2019darwin} using host-side code modifications.
    
    5. \textbf{Traceback Pointer Data Types and States:} \label{sec:traceback_states}  Traceback pointers in DP-HLS are stored in data type \texttt{tb\_t} which users may define using arbitrary precision data type provided in Vitis HLS. For the Global Linear kernel (\hyperref[kernel:1]{\#1}), \texttt{tb\_t} is defined as \texttt{ap\_uint<2>}  and for the Global Affine kernel (\hyperref[kernel:2]{\#2}), as \texttt{ap\_uint<4>}, since the recurrence equations require a minimum of 2-bits and 4-bits, respectively, to represent the traceback pointers. 
    
    The traceback logic in the final step of DP algorithms is equivalent to a finite state machine (FSM) in which the current state and DP matrix cell score determine the next state and cell in the score matrix, which is translated to the traceback path. In DP-HLS, users enumerate the possible traceback states in the variable \texttt{TB\_STATE}. For example, the Global Affine kernel (\hyperref[kernel:2]{\#2}) enumerates three states — \texttt{MM}, \texttt{INS}, and \texttt{DEL} — representing match/mismatch, insertion, and deletion (Listing \ref{listing:traceback} (left)). In Listing \ref{listing:traceback} (right), the Global Two-piece Affine kernel (\hyperref[kernel:5]{\#5}) has two extra traceback states (\texttt{LONG\_INS, LONG\_DEL}), one for each additional recurrence equation modeling long gap scores. The framework offers a no-traceback option, used by the Viterbi (\hyperref[kernel:10]{\#10}) and Banded Local Affine (\hyperref[kernel:12]{\#12}) kernels, to skip the traceback step.
    
\begin{figure}
\captionsetup{type=lstlisting}
\begin{minipage}{.20\textwidth}
\begin{lstlisting}[language=C++, numbers=none]
enum TB_STATE {
    MM, INS, DEL
} tb_next_state, tb_curr_state;
\end{lstlisting}
\end{minipage}
\begin{minipage}{.22\textwidth}
\begin{lstlisting}[language=C++, numbers=none]
enum TB_STATE {
    MM, INS, DEL, 
    LONG_INS, LONG_DEL 
} tb_next_state, tb_curr_state;
\end{lstlisting}
\end{minipage}
\caption{Definition of traceback states for \textbf{(Left)} Global Affine (\hyperref[kernel:2]{\#2}) and \textbf{(Right)} Global Two-piece Affine (\hyperref[kernel:5]{\#5}) kernels.}
\label{listing:traceback}
\end{figure}

    6. \textbf{Banding Width:} DP-HLS allows users to use a fixed banding search space pruning strategy by setting the macros \texttt{BANDING} and \texttt{BANDWIDTH} to the desired band size.
    
\noindent \textbf{{\encircle{2}} Initializing Row and Column Scores:} The DP-HLS framework includes two internal 2-D arrays, \texttt{init\_row\_scr} and \texttt{init\_col\_scr},  to store the scores of the initial row and column of the DP matrix, with the dimension of \texttt{MAX\_REFERENCE\_LENGTH$\times$N\_LAYERS} and \texttt{MAX\_QUERY\_LENGTH$\times$N\_LAYERS}, respectively. The users should specify the values of these arrays, as defining how DP-HLS would initialize them at the runtime. Listing \ref{listing:initialization} depicts the initialization step of the Global Linear kernel (\hyperref[kernel:1]{\#1}), which has a single scoring layer at index 0 whose first row and column are initialized to account for gaps at the start of the alignment.

\begin{figure}
\captionsetup{type=lstlisting}
\begin{lstlisting}[language=C++, numbers=none]
type_t gap = scoring_params.linear_gap;
for (int i = 0; i < MAX_REFERENCE_LENGTH; i++){
    init_row_scr[i][0] = i*gap; }
for (int i = 0; i < MAX_QUERY_LENGTH; i++){
    init_col_scr[i][0] = i*gap; }
\end{lstlisting}
\caption{Initialization of column and row scores for Global Linear kernel (\hyperref[kernel:1]{\#1}).}
\label{listing:initialization}
\end{figure}

\noindent \textbf{{\encircle{3}} Specifying PE Function:} In DP-HLS, all computations involved in the \emph{matrix fill} step of the 2-D DP kernels (Section \ref{sec:scoring}) are executed by computing units known as \emph{processing elements} (PEs). Within the DP-HLS front-end component, users only need to specify the recurrence equations for computing the score and traceback pointer for a single cell (\texttt{i}, \texttt{j}), located at row \texttt{i} and column \texttt{j} of the DP matrix. The back-end component automatically manages systolic communication between PEs and the storage of data in memory buffers. The recurrence equations are specified in \texttt{PE\_func} (Listing \ref{listing:pe_function}) For example, in the Local Linear kernel (\hyperref[kernel:3]{\#3}), the arrays \texttt{dp\_mem\_up}, \texttt{dp\_mem\_diag}, and \texttt{dp\_mem\_left}, are the inputs to \texttt{PE\_func} and populated with cell scores by the DP-HLS back-end for cells at positions up (\texttt{i-1}, \texttt{j}), diagonal (\texttt{i-1}, \texttt{j-1}) and left (\texttt{i}, \texttt{j-1}) of the current cell (\texttt{i}, \texttt{j}). Likewise, the \texttt{i$^{th}$} query character and the \texttt{j$^{th}$} reference character are also automatically available to the input of \texttt{PE\_func} as \texttt{lc\_qry\_val} and \texttt{lc\_ref\_val}, respectively. At the end of the function call, valid scores and traceback pointers for cell (\texttt{i}, \texttt{j}) must be stored to \texttt{wt\_scr} and \texttt{wt\_tbp} (Listing~\ref{listing:max_traceback}). 

\begin{figure}
\captionsetup{type=lstlisting}
\begin{lstlisting}[language=C++, numbers=none]
type_t linear_gap = params.linear_gap;
type_t ins = dp_mem_left[0] + linear_gap;
type_t del = dp_mem_up[0] + linear_gap;
type_t match = dp_mem_diag[0] + (lc_qry_val == lc_ref_val) ? params.match : params.mismatch;
\end{lstlisting}
\caption{Definition of \texttt{PE\_func} for Local Linear kernel (\hyperref[kernel:3]{\#3}) computing the scores of the DP matrix.}
\label{listing:pe_function}
\end{figure}

\begin{figure}
\captionsetup{type=lstlisting}
\begin{lstlisting}[language=C++, numbers=none]
type_t max_value = ins;
wt_tbp = TB_LEFT;
if (max_value < match){ max_value = match;
    wt_tbp = TB_DIAG; }
if (max_value < del){ max_value = del;
    wt_tbp = TB_UP; }
if (max_value < (type_t)0) { max_value = 0;
    wt_tbp = TB_END; }
wt_scr = max_value; 
\end{lstlisting}
\caption{Logic for finding max cell score in the DP matrix and traceback pointer for Local Linear kernel (\hyperref[kernel:3]{\#3}).}
\label{listing:max_traceback}
\end{figure}

\noindent \textbf{{\encircle{4}} Specifying Traceback Strategy:} 
An FSM intuitively defines the traceback logic in the final stage of the DP algorithm (Section \ref{sec:traceback_strategy}). Section \hyperref[sec:traceback_states]{4.1.5} mentions the customization of the FSM states and traceback pointers. For multiple scoring matrices, each matrix maps to a state, and transitions represent jumps between matrices. Listing \ref{listing:traceback_mapping} defines the logic to map the current state and pointer to the next cell at each traceback step. In the Local Linear kernel (\hyperref[kernel:3]{\#3}), a single state example, an outer if-statement checks the current \texttt{tb\_state}, assigns the new state, and directs the traceback via \texttt{wt\_tbp} for \texttt{INS}, \texttt{DEL}, \texttt{MM}, or \texttt{AL\_END} (end of traceback).

\begin{figure}
\captionsetup{type=lstlisting}
\begin{lstlisting}[language=C++, numbers=none]
if (tb_state == TB_STATE::MM){
    if (tb_ptr == TB_DIAG){tb_move = AL_MMI; }
    else if (tb_ptr == TB_UP){tb_move = AL_DEL; }
    else if (tb_ptr == TB_LEFT){tb_move = AL_INS;}
    else if (tb_ptr == TB_END) {tb_move = AL_END;}
    else {tb_move = AL_END;}
    tb_state = TB_STATE::MM;}
\end{lstlisting}
\caption{Definition of traceback logic state transitions for Local Linear kernel (\hyperref[kernel:3]{\#3}).}
\label{listing:traceback_mapping}
\end{figure}

\noindent \textbf{{\encircle{5}} Specifying Parallelism:} 
The front-end component provides parameters (\texttt{N\textsubscript{PE}}, \texttt{N\textsubscript{B}}, \texttt{N\textsubscript{K}}) that users can adjust to parallelize the synthesized design empirically, without understanding the internal details of the DP-HLS back-end and compiler optimizations. \texttt{N\textsubscript{PE}} determines the level of \textit{inner-loop parallelism} for a single pair of sequences. DP-HLS also exploits \textit{outer-loop parallelism} across multiple sequence pairs by setting the parameters \texttt{N\textsubscript{B}} and \texttt{N\textsubscript{K}}. This allows the processing of \texttt{N\textsubscript{B} $\times$ N\textsubscript{K}} independent sequence pairs, with \texttt{N\textsubscript{K}} independent channels to the host CPU (to take advantage of CPU multi-threading, for example), each consisting of \texttt{N\textsubscript{B}} blocks sharing a single arbiter on the device, as shown in Fig. \ref{fig:implementation}B. The design allows linking \texttt{N\textsubscript{K}} heterogeneous kernels (e.g., a mix of global and local aligners) seamlessly in the design, a process that would be quite cumbersome with HDL.

\noindent \textbf{{\encircle{6}} Writing Host-Side Program:} After completing the device specification, the user must define the host application to manage pre-processing and transfer of input sequences to the device, invoke device kernels, and receive alignment output from the device and/or perform tiling (for long reads). Effective scheduling is important to optimize device utilization; therefore, for optimal performance, it should be designed to schedule batches of input sequences and use multi-threading to leverage the device's \texttt{N\textsubscript{K}} independent channels. The host program uses OpenCL syntax \cite{munshi2011opencl}, and relevant examples from our 15 implemented kernels (Table \ref{table:summary}) are provided in the DP-HLS GitHub repository.

\section{Back-end Implementation}
\label{sec:back-end}



The DP-HLS back-end component is based on the efficient mapping of 2-D DP algorithms to a linear systolic array, as extensively used in previous hardware accelerators \cite{fei2018fpgasw, turakhia2018darwin, lindegger2023scrooge,turakhia2019darwin, haghi2021fpga, zhang2007implementation, msa_dp, bsw_systolic, seedex, dp_2_piece, dp_protein_affine, bwamem_accel, msa_fpga}. This enables various HLS optimizations to be encapsulated together without user modifications, thereby adding an abstraction layer to the HLS design, which helps boost productivity. DP-HLS back-end also imposes design constraints on scoring and traceback stages to ensure optimal performance and configurability. Fig.~\ref{fig:implementation}B-C depicts the key back-end optimizations in DP-HLS using labels \textbf{\encircle{a}}-\textbf{\encircle{i}}. Optimizations \textbf{\encircle{a}}-\textbf{\encircle{f}} serve as hints to the HLS compiler to produce a linear systolic array of PEs that exploits \emph{wavefront (anti-diagonal) parallelism} while computing the scores of the DP matrix, whereas optimizations \textbf{\encircle{g}}-\textbf{\encircle{i}} apply to the traceback logic and for exploiting inter-task parallelism. These optimizations are elaborated below.



\subsection{Scoring Logic Design}
\label{subsec:scoring}
Fig.~\ref{fig:implementation}C illustrates the scoring operation using an example of linear systolic array of 
four PEs (\texttt{N\textsubscript{PE}}=4), which are used to compute the DP matrix. The rows of the DP matrix are divided into \emph{chunks}, with each chunk containing \texttt{N\textsubscript{PE}} 
consecutive rows, each assigned to a different PE (each row is color-coded according to the PE responsible for its computation). During the processing of a query chunk, each PE is initialized with the query base-pair corresponding to its row in the chunk while the reference sequence streams through the array. The systolic array architecture exploits the wavefront parallelism inherent in the computation of the query chunk. Chunk-wise score computation requires a buffer (\texttt{Preserved Row Score Buffer}) to store the scores computed by the last PE, which are then used by the first PE of next query chunk. The buffer size corresponds to the maximum length of the reference sequence in the DP matrix.

In the back-end component, the 2-D DP matrix is computed using the user-defined \texttt{PE\_func} within nested \texttt{for} loops. The \textit{outer loop} divides the matrix into chunks, the \textit{middle loop} iterates through the wavefronts in this chunk, and the \textit{inner loop} unrolls the PE function to map each score in the wavefront to its corresponding PE. Within the middle loop, the back-end applies optimization \textbf{\encircle{a}} using the \texttt{\#pragma HLS PIPELINE} directive to enable wavefront pipelining. Each stage of the pipeline starts the computation of a new wavefront of the DP matrix using the unrolled PE array, while each PE in the array calls \texttt{PE\_func} at the same cycle. The number of clock cycles between the initialization of each wavefront can be controlled by setting the \textit{Initialization Interval (II)} of this directive, with \texttt{II=1} as the optimum. For complex PE functions requiring more than 1 cycle per wavefront, HLS finds the minimum possible \texttt{II} to meet the timing requirement. At the inner loop, we apply optimization \textbf{\encircle{b}} using the \texttt{\#pragma HLS UNROLL} directive to create a linear systolic array of PEs. However, to correctly and efficiently unroll the PEs, each input and output data element must be parallelly accessible for each PE. To ensure this, DP-HLS creates several fully partitioned local buffers that allow parallel data access using the \texttt{\#pragma HLS ARRAY PARTITION variable=<buffer> type=complete} directive. For example, optimizations \textbf{\encircle{c}} and \textbf{\encircle{d}} apply this directive to create local reference and query buffers, respectively, which store the sequence characters at the current wavefront. This directive is also used in the optimization \textbf{\encircle{e}} for \texttt{DP Memory Buffer} and \texttt{Score Buffer}, which stores the previous two wavefronts and the output scores of the current wavefront.

\subsection{Traceback Logic and Memory Design}
The optimal design of traceback logic and memory is essential for efficiency, as it typically consumes the most amount of memory resources in 2-D DP algorithms that require traceback. Below, we describe the two most critical optimizations implemented by the DP-HLS back-end component.

\emph{First}, the back-end reorganizes the 2-D DP matrix so that the first dimension corresponds to \texttt{N\textsubscript{PE}}, while the second dimension is scaled to accommodate the total pointers for user-specified maximum reference and query lengths, i.e., \texttt{MAX\_REFERENCE\_LENGTH} and \texttt{MAX\_QUERY\_LENGTH}, respectively. This reorganization ensures that each PE has access to a dedicated memory bank, allowing it to independently store its traceback (TB) pointers every cycle, thereby minimizing the \texttt{II} of the wavefront loop. The back-end further optimizes memory access by applying address coalescing, which translates consecutive wavefronts in the DP matrix to consecutive columns in the TB memory (see \textbf{{\encircle{g}}} in Fig.~\ref{fig:implementation}). This results in a more regular and efficient access pattern as all the PEs write their TB pointers to the same address in different memory banks. Additionally, the back-end automatically keeps track of the corresponding addresses each wavefront shall write in the TB memory.

\emph{Second}, the back-end includes logic that allows users to configure the start and end conditions of the traceback. For instance, some traceback strategies require locating the maximum scores in the last row or column of the DP matrix. In such cases, each PE tracks its local maximum of the scores it computes if its coordinate satisfies the requirements (at the last row or column). Then, the DP-HLS back-end incorporates reduction logic to identify the maximum cell of the entire DP matrix block within a few cycles by a reduction maximum over each PE's local max. The reduction operations happen once for each alignment block to determine the global maximum scoring cell.

\subsection{Parallel Execution of Kernels}

As described in Section~\ref{sec:front-end}, users specify parallelism using three parameters: \texttt{N\textsubscript{K}}, \texttt{N\textsubscript{B}}, and \texttt{N\textsubscript{PE}}. The parallelism for \texttt{N\textsubscript{PE}} is managed by the scoring logic optimizations discussed in Section~\ref{subsec:scoring}, while \texttt{N\textsubscript{K}} is handled by the linker. To enable parallel processing of \texttt{N\textsubscript{B}} blocks within a kernel, DP-HLS uses two main optimizations. \textit{First}, it uses \texttt{\#pragma HLS UNROLL} to create multiple blocks within a kernel, allowing for the concurrent execution of these blocks (see \textbf{{\encircle{h}}} in Fig.~\ref{fig:implementation}). \textit{Second}, it ensures concurrent memory access to the input and output buffers for each block through block-wise partition, using \texttt{\#pragma HLS ARRAY\_PARTITION type=block dim=1}, with the number of partitions set to \texttt{N\textsubscript{B}} (see \textbf{{\encircle{i}}} in Fig.~\ref{fig:implementation}).

\section{Methodology}
\label{sec:method}

\subsection{Datasets}
    \noindent\textbf{DNA Sequences:} The DNA sequences were generated using PBSIM2 \cite{ono_pbsim2_2021} by simulating 1,000 PacBio \cite{rhoads_pacbio_2015} reads of 10,000 bases with a 30\% error rate from the human reference genome GRCh38. For tiling version of Kernel \hyperref[kernel:2]{\#2}, full reads were used. For short alignment kernels (\hyperref[kernel:1]{\#1}-\hyperref[kernel:7]{7}, \hyperref[kernel:10]{\#10}-\hyperref[kernel:13]{13}), each read was truncated to 256 bases. 
    
    \noindent\textbf{Protein Sequences: } Protein sequences for Kernel \hyperref[kernel:15]{\#15} were randomly sampled from UniProtKB (we used Swiss-Prot, \texttt{2024\_03} release version, containing 571k entries) \cite{the_uniprot_consortium_uniprot_2023}. 

    
    \noindent\textbf{Complex Number Sequences: } We simulated our own sequences using randomly generated complex numbers for input to the DTW kernel (\hyperref[kernel:9]{\#9}).

   \noindent \textbf{sDTW Sequences: } For the sDTW kernel (\hyperref[kernel:14]{\#14}), we used the datasets used in the SquiggleFilter implementation \cite{dunn2021squigglefilter}.

   \noindent \textbf{Sequence Profiles: } For the Profile Alignment kernel (\hyperref[kernel:8]{\#8}), we generated profiles from randomly selected regions of  256 base pairs across the genomic sequences of \textit{Drosophila melanogaster} and \textit{Drosophila simulans}.


\subsection{DP-HLS Framework Evaluation}

\noindent\textbf{Implementation:} We implemented 15 DP-HLS kernels using \texttt{C++(v17)} (Table ~\ref{table:summary}) and performed a \texttt{C++}-based simulation to verify the correctness of the final alignment output. The kernels are then synthesized using AMD\textsuperscript{\textregistered} Vitis HLS \texttt{2021.2} tool \cite{VitisHighLevelSynthesis2021}, present on AWS\textsuperscript{\textregistered} FPGA Developer AMI (\texttt{v1.12.2}), by following the complete design flow – from \texttt{C/C++}-Simulation to Bitstream Generation (Fig.~\ref{fig:implementation}A), and deployed on AWS\textsuperscript{\textregistered} EC2 F1 instance (\texttt{f1.2xlarge}; with FPGA device \texttt{XCVU9P-FLGB2104-2-I}) using \texttt{v++} command-line tools and OpenCL-based host code for host-kernel communication. For Kernel \hyperref[kernel:2]{\#2}, we applied host-side code changes to support long alignments using a tiling heuristic \cite{turakhia2018darwin}. A fixed target frequency of 250 MHz was set before synthesis. Parameters \texttt{N\textsubscript{PE}}, \texttt{N\textsubscript{B}} and \texttt{N\textsubscript{K}} were configured for each kernel to maximize the device throughput  (Table \ref{table:max_throughput}). 


\noindent\textbf{Throughput:} Throughput for each DP-HLS kernel was calculated by using the number of clock cycles reported in the \textit{co-simulation} step of AMD\textsuperscript{\textregistered} Vitis HLS tool, maximum achievable frequency, and the number of parallel alignments computed on the FPGA device.  

\noindent\textbf{Resource Utilization:} Post-routing reports of the bitstream generation step are used to obtain the Block RAMs (\textbf{BRAM}), Flip-Flops (\textbf{FF}), Lookup Tables (\textbf{LUT}), and Digital Signal Processors (\textbf{DSP}) utilization on the FPGA device for all kernels 
 reported as a percentage of the total resources available on the AWS\textsuperscript{\textregistered} EC2 F1 FPGA device. 

\noindent\textbf{Scalability:} To examine the scalability of the kernels, we chose two diverse kernels, Global Linear (\hyperref[kernel:1]{\#1}) and DTW (\hyperref[kernel:9]{\#9}), and evaluated their resource utilization and throughput values with increasing \texttt{N\textsubscript{PE}} and \texttt{N\textsubscript{B}} and fixed operating frequency of 250 and 200 MHz (Section \ref{subsec:throughput_scaling}).

\subsection{Baseline Comparison}

\noindent\textbf{Software Baselines:} 
\label{sec:methods_software}
We compared the throughput of DP-HLS Kernels \hyperref[kernel:1]{\#1}-\hyperref[kernel:7]{7}, \hyperref[kernel:11]{\#11}-\hyperref[kernel:12]{12}, \hyperref[kernel:15]{\#15} with state-of-the-art parallel CPU implementations, using SeqAn3~\cite{reinertSeqAnTemplateLibrary2017} (\texttt{v3.3.0}), a widely-used, multi-threaded bioinformatics library, as the baseline. For the Two-Piece affine (\hyperref[kernel:5]{\#5}) and protein sequence alignment (\hyperref[kernel:15]{\#15}) kernels, we used Minimap2 \cite{li_minimap2_2018} (\texttt{v2.28}) and the command-line version of EMBOSS Water \cite{PMID:38597606} (\texttt{v6.6.0}) as our software baselines, respectively. All software baselines were tested in a system with \texttt{g++(v10.3)} and \texttt{cmake(v3.16.3)} and executed on a 36-core CPU-optimized AWS\textsuperscript{\textregistered} EC2 instance, \texttt{c4.8xlarge} with 60 GB memory, which costs \$1.591 per hour, comparable to the AWS\textsuperscript{\textregistered} EC2 F1 instance, \texttt{f1.2xlarge} (\$1.650 per hour) on which the DP-HLS throughput is estimated. The throughput values for SeqAn3 and Minimap2 were calculated by setting number of threads 
as 32 and considering the wall clock time of total execution. Since EMBOSS lacks multi-threading, we measured its throughput of 32 parallel jobs launched with GNU \texttt{parallel}.

\noindent\textbf{Hardware Baselines: }
\label{sec:methods_hardware}
We compared the FPGA throughput and resource utilization of the DP-HLS kernels with hand-crafted RTL implementations (\textit{RTL Baselines}), GPU implementations (\textit{GPU Baselines}), and previous HLS implementations (\textit{HLS baselines}). 

\textit{RTL Baselines}: We obtained optimized open-source RTL implementations of GACT \cite{turakhia2018darwin},  Banded Smith-Waterman (BSW) \cite{turakhia2019darwin} (\texttt{v1}) and SquiggleFilter \cite{dunn2021squigglefilter} (\texttt{v1.1.0}) accelerators as RTL baselines to compare with Kernels \hyperref[kernel:2]{\#2}, \hyperref[kernel:12]{\#12}, and \hyperref[kernel:14]{\#14} (Table \ref{table:summary}), respectively, since all are based on linear systolic array architecture, similar to DP-HLS kernels. The baselines are implemented using AMD\textsuperscript{\textregistered} Vivado \texttt{2021.2} tool on the same AWS\textsuperscript{\textregistered} EC2 F1 FPGA instance (\texttt{f1.2xlarge}), and resource utilization numbers were collected from the \textit{Implementation} step. Throughput values for the first two were measured via Icarus Verilog simulations and BSW kernel via Vivado waveform simulations. To demonstrate the scaling effects, DP-HLS's Kernel \hyperref[kernel:2]{\#2} was compared with GACT with increasing \texttt{N\textsubscript{PE}}.  The match-bonus feature in SquiggleFilter was removed to match Kernel \hyperref[kernel:14]{\#14}’s implementation.




%



\textit{GPU Baselines}: CUDASW++\texttt{4.0} \cite{schmidt2023cudasw++}, which provides GPU implementation of the Smith-Waterman algorithm for protein sequences, was used as the baseline for DP-HLS Kernel \hyperref[kernel:15]{\#15}. We disabled the traceback step in DP-HLS since it is not performed in the baseline. We used GASAL2 \cite{ahmed2019gasal2}, with \texttt{alignment type} set as \texttt{LOCAL}, \texttt{GLOBAL}, \texttt{BSW}, as baselines for DP-HLS Kernels \hyperref[kernel:4]{\#4}, \hyperref[kernel:2]{\#2}, and \hyperref[kernel:12]{\#12}, respectively. 
Both baselines were tested on an AWS\textsuperscript{\textregistered} EC2 \texttt{p3.2xlarge} instance with NVIDIA\textsuperscript{\textregistered} Tesla V100 GPU (costing \$3.06/hour). 
All throughput values were normalized for iso-cost comparison with the AWS\textsuperscript{\textregistered} EC2 F1 FPGA instance (costing \$1.65/hour). 

\textit{HLS Baselines}: As HLS baselines, we used the AMD\textsuperscript{\textregistered} Vitis Genomics Library, containing optimized Vitis HLS Libraries \cite{xilinx_vitis_libraries}. We used the \texttt{2021.2} branch, which works with the Vitis HLS \texttt{2021.2} version used for DP-HLS implementation. The Smith-Waterman kernel in this library matches with DP-HLS Kernel \hyperref[kernel:3]{\#3} implementation, so we chose it as our HLS baseline, with \texttt{N\textsubscript{PE}=32, N\textsubscript{K}=1, N\textsubscript{B}=32} and the maximum target clock frequency of 333 MHz.

\section{Results}
\label{sec:results}

\subsection{DP-HLS efficiently implements diverse 2-D DP kernels}

\renewcommand{\tabcolsep}{1pt}
\begin{table}
\tiny
\centering
\begin{tabularx}{0.4515\textwidth}{|c|c|c|c|c|c|c|c|}
  \hline
  {\textbf{Kernel}} &
      \multicolumn{4}{c|}{\textbf{Resource utilization for a 32PE block}} &
      {\textbf{Optimal}} &
      {\textbf{Max}} &
      {\textbf{Alignments/}} \\
     \textbf{No.} & \textbf{LUT} & \textbf{FF} & \textbf{BRAM} & \textbf{DSP} &  (\texttt{N\textsubscript{PE},N\textsubscript{B}, N\textsubscript{K}}) & \textbf{Freq} (MHz) & \textbf{sec}\\ 
  \hline
  \hyperref[kernel:1]{\#1} & 0.72\% & 0.42\% & 1.78\% & 0.029\% & (64,16,4) & 250.0 & $3.51 \times 10^6$ \\
  \hline
  \hyperref[kernel:2]{\#2} & 1.30\% & 0.517\% & 1.78\% & 0.029\% & (32,16,4) & 250.0 & $2.85 \times 10^6$ \\
  \hline
  \hyperref[kernel:3]{\#3} & 0.95\% & 0.63\% & 1.67\% & 0.014\% & (32,16,5) & 250.0 & $3.43 \times 10^6$ \\
  \hline
  \hyperref[kernel:4]{\#4} & 1.60\% & 0.75\% & 1.67\% & 0.014\% & (32,16,4) & 250.0 & $2.71 \times 10^6$\\
  \hline
  \hyperref[kernel:5]{\#5} & 2.03\% & 0.65\% & 2.67\% & 0.029\% & (32,8,5) & 150.0 & $1.06 \times 10^6$ \\
  \hline
  \hyperref[kernel:6]{\#6} & 0.98\% & 0.66\% & 1.67\% & 0.014\% & (32,16,4) & 250.0 & $2.73 \times 10^6$\\
  \hline
  \hyperref[kernel:7]{\#7} & 1.17\% & 0.67\% & 0.83\% & 0.014\% & (32,16,4) & 250.0 & $3.34 \times 10^6$ \\
  \hline
  \hyperref[kernel:8]{\#8} & 3.66\% & 2.56\% & 2.56\% & 28.11\% & (16,1,5) & 166.7 & $3.70 \times 10^4$\\
  \hline
  \hyperref[kernel:9]{\#9} & 1.62\% & 1.55\% & 1.88\% & 2.84\% & (64,4,3) & 200.0 & $2.31 \times 10^5$ \\
  \hline
  \hyperref[kernel:10]{\#10} & 3.78\% & 1.69\% & 1.67\% & 0.014\% & (16,4,7) & 125.0 & $4.90 \times 10^5$ \\
  \hline
  \hyperref[kernel:11]{\#11} & 1.02\% & 0.40\% & 0.94\% & 0.029\% & (64,8,7) & 166.7 & $2.25 \times 10^6$ \\
  \hline
  \hyperref[kernel:12]{\#12} & 1.44\% & 0.70\% & 0.57\% & 0.014\% & (16,16,7) & 200.0 & $4.77 \times 10^6$ \\
  \hline
  \hyperref[kernel:13]{\#13} & 2.25\% & 0.69\% & 1.83\% & 0.029\% & (16,8,7) & 125.0 & $1.24 \times 10^6$ \\
  \hline
  \hyperref[kernel:14]{\#14} & 1.22\% & 0.76\% & 0.57\% & 0.014\% & (32,16,5) & 250.0 & $5.16 \times 10^6$ \\
  \hline
  \hyperref[kernel:15]{\#15} & 1.47\% & 0.95\% & 2.56\% & 0.014\% & (32,8,5) & 200.0 & $9.33 \times 10^5$ \\
  \hline
\end{tabularx}
\caption{\textbf{Performance summary of 15 DP-HLS Kernels.} The kernel numbers correspond to the kernels listed in Table \ref{table:summary}. Utilization in $\%$ of available FPGA resources is shown for a single block for 32 PEs for uniformity.  We also show the optimal configuration of (\texttt{N\textsubscript{PE}, N\textsubscript{B}, N\textsubscript{K}}) for each kernel that resulted in maximum throughput (alignments per second), and its corresponding maximum clock frequency and throughput achieved.} 
\label{table:max_throughput}
\end{table}

Table \ref{table:max_throughput} summarizes the performance and resource utilization of all 15 2-D DP kernels implemented on DP-HLS, as listed in Table \ref{table:summary}. These kernels vary widely in terms of applications and computational patterns (Table \ref{table:summary}), which is also evident from the range of hardware resource utilization and throughput values presented in Table \ref{table:max_throughput}.

For instance, when comparing the resource utilization of a single block of 32 PEs, Profile Alignment (\hyperref[kernel:8]{\#8}) and DTW (\hyperref[kernel:9]{\#9}) kernels have relatively higher DSP consumption. This is expected as these kernels perform multiplications, with Kernel \#8 requiring two matrix-vector multiplications in each cell. For most other kernels, the DSP is only used for pre-computing traceback addresses.

The BRAM usage is primarily influenced by the complexity of traceback across most kernels. For instance, Kernels \hyperref[kernel:5]{\#5} and \hyperref[kernel:13]{\#13}, which implement a two-piece affine gap penalty, require more BRAM as they need at least 7 bits to store each pointer, compared to only 2 bits per pointer for kernels with a linear penalty (e.g., Kernels \hyperref[kernel:1]{\#1} and \hyperref[kernel:3]{\#3}). In Kernel \hyperref[kernel:10]{\#10} and \hyperref[kernel:12]{\#12}, BRAM usage is minimal since traceback is not involved. Kernel \hyperref[kernel:15]{\#15}, which deals with protein alignment, also consumes more BRAM to store the larger substitution matrix in \texttt{ScoringParams} (20$\times$20 for protein sequences, compared to 4$\times$4 for certain alignment algorithms of DNA sequences).



LUT and FF usage is mainly influenced by the complexity of the scoring equations. For instance, Kernel \hyperref[kernel:8]{\#8}, with its complex matrix-vector multiplication, shows the highest FF and LUT utilization, followed closely by the Viterbi kernel (\hyperref[kernel:10]{\#10}). Banding kernels (\hyperref[kernel:11]{\#11}-\hyperref[kernel:13]{13}) also have slightly elevated logic usage due to the extra computations needed to determine the bands. 

In terms of throughput, resource-intensive kernels (\hyperref[kernel:8]{\#8}-\hyperref[kernel:10]{10}) have relatively lower values due to their complex computational patterns. For instance, Kernel \hyperref[kernel:8]{\#8} requires multiple cycles (\texttt{II=4}) to compute a single DP cell due to matrix-vector multiplications. The complexity of the scoring equations also impacts clock frequency, as seen in the Viterbi (\hyperref[kernel:10]{\#10}) and Banded Global Two-piece Affine (\hyperref[kernel:13]{\#13}) kernels. Overall, the resource utilization, clock frequency, and throughput numbers suggest that all kernels have been efficiently implemented by DP-HLS, which is also confirmed by the baseline comparisons presented later. 

\subsection{DP-HLS kernel implementations demonstrate 1-D systolic array behavior}

\label{subsec:throughput_scaling}

\begin{figure}
\tiny
    \includegraphics[width=0.47\textwidth]{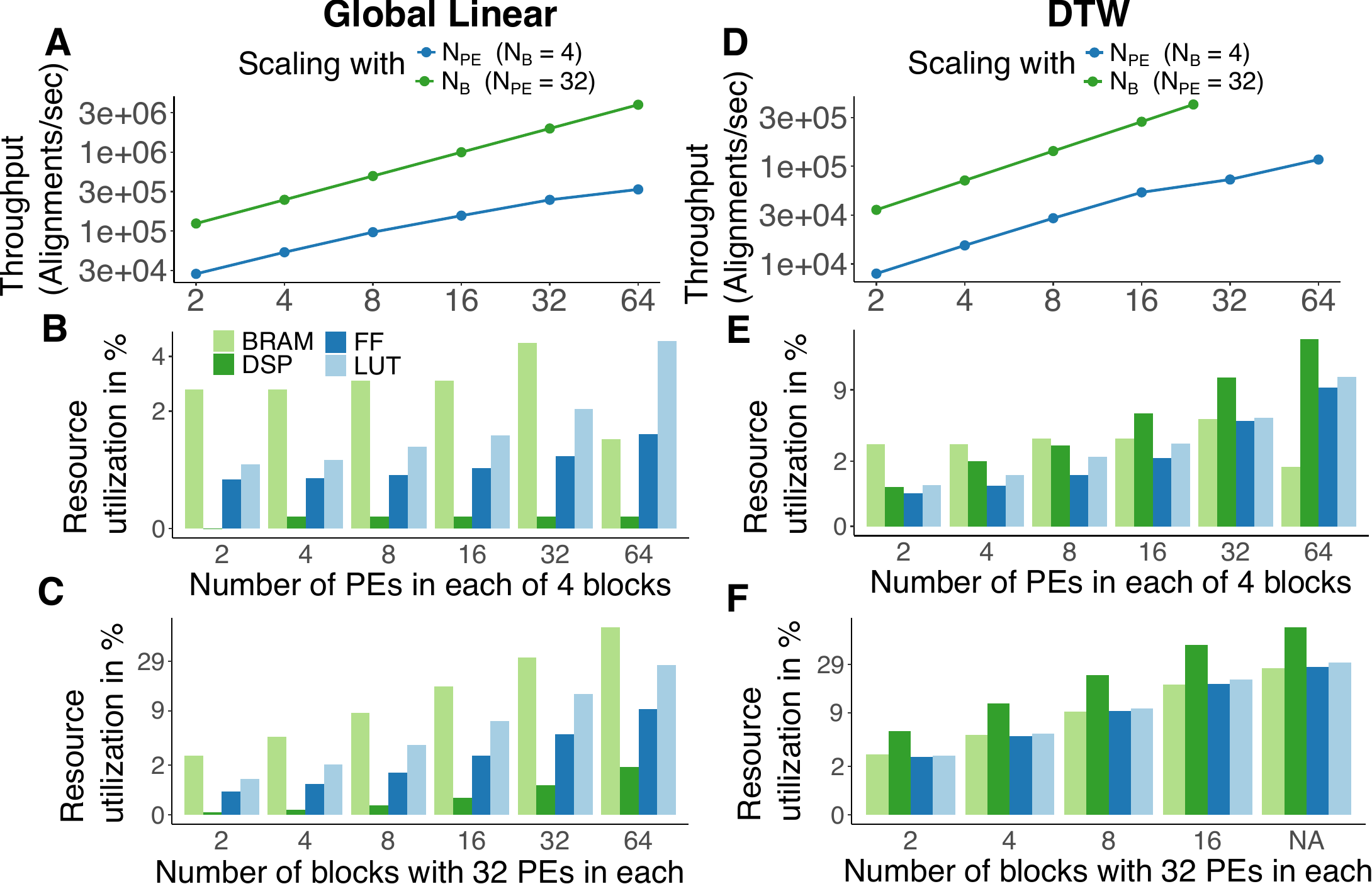}
    \caption{\textbf{Scaling results for Global Linear (\hyperref[kernel:1]{\#1}) and DTW (\hyperref[kernel:9]{\#9}) kernels with  \texttt{N\textsubscript{PE}} and \texttt{N\textsubscript{B}}}. (\textbf{A, D}) Throughput scaling of Global Linear (A) and DTW (D) in \texttt{log-log} scale with \texttt{N\textsubscript{PE}} and \texttt{N\textsubscript{B}}; (\textbf{B-C}) Resource utilization scaling for Global Linear kernel; (\textbf{E-F}) Resource utilization scaling for DTW kernel.}
    \label{fig:resources}
\end{figure}

Through the DP-HLS back-end optimizations, we would expect the HLS compiler to produce \texttt{N\textsubscript{B}} 1-D systolic arrays, each with \texttt{N\textsubscript{PE}} processing elements. However, since HLS-generated code is not easily interpretable, it is difficult to verify this directly from the code. Hence, we varied \texttt{N\textsubscript{PE}} and \texttt{N\textsubscript{B}} and checked if the throughput and resources scaled according to the expected behavior. Fig. \ref{fig:resources} presents these results for two diverse kernels, Global Linear (\hyperref[kernel:1]{\#1}) and DTW (\hyperref[kernel:9]{\#9}), but we observed similar patterns across all 15 kernels.



Fig. \ref{fig:resources}A,D shows that throughput scales nearly perfectly with \texttt{N\textsubscript{PE}} at lower values for both kernels but experiences some saturation at higher values. This is expected as the wavefront parallelism exploited by systolic arrays diminishes near the edges of the DP matrix, leading to more PEs with more idle cycles. In contrast, the large inter-alignment parallelism exploited by the \texttt{N\textsubscript{B}} independent arrays allows throughput to scale almost perfectly with \texttt{N\textsubscript{B}}, which is confirmed by the results for both kernels in Fig. \ref{fig:resources}. For  DTW, \texttt{N\textsubscript{B}} is capped at 24 as it reached maximum DSP availability.

Also, as we would expect, the percentage utilization of all resource types scales almost perfectly with increasing \texttt{N\textsubscript{B}} while keeping \texttt{N\textsubscript{PE}} constant (Fig. \ref{fig:resources}C, F). This occurs because each parallel block is identical, leading to a proportional increase in resource usage. When \texttt{N\textsubscript{B}} is fixed, LUT and FF utilization scales perfectly with increasing \texttt{N\textsubscript{PE}} for both kernels (Fig. \ref{fig:resources}B, E) due to the linear systolic array structure. However, DSP usage varies depending on the kernel’s algorithm. For instance, DSP utilization scales well with increasing \texttt{N\textsubscript{PE}} in the DTW kernel (Fig. \ref{fig:resources}E) as DSPs are used by each PE for the scoring logic, whereas for the Global Linear kernel, DSP usage remains constant (Fig. \ref{fig:resources}B) since DSPs are used for fixed logic outside the PEs to precompute the traceback starting address. Additionally, BRAM is primarily used for TB memory, which increases with \texttt{N\textsubscript{PE}} up to 32 but does not scale proportionally. For higher \texttt{N\textsubscript{PE}} values (e.g., \texttt{N\textsubscript{PE}=64}), HLS compiler optimizations sometimes convert BRAMs to LUTRAMs to reduce memory access latency, resulting in lower BRAM usage.

\subsection{DP-HLS kernels provide competitive performance to optimized RTL implementations}
\label{sec:hardware_baseline_result}

\begin{figure}
    \includegraphics[width=0.44\textwidth]{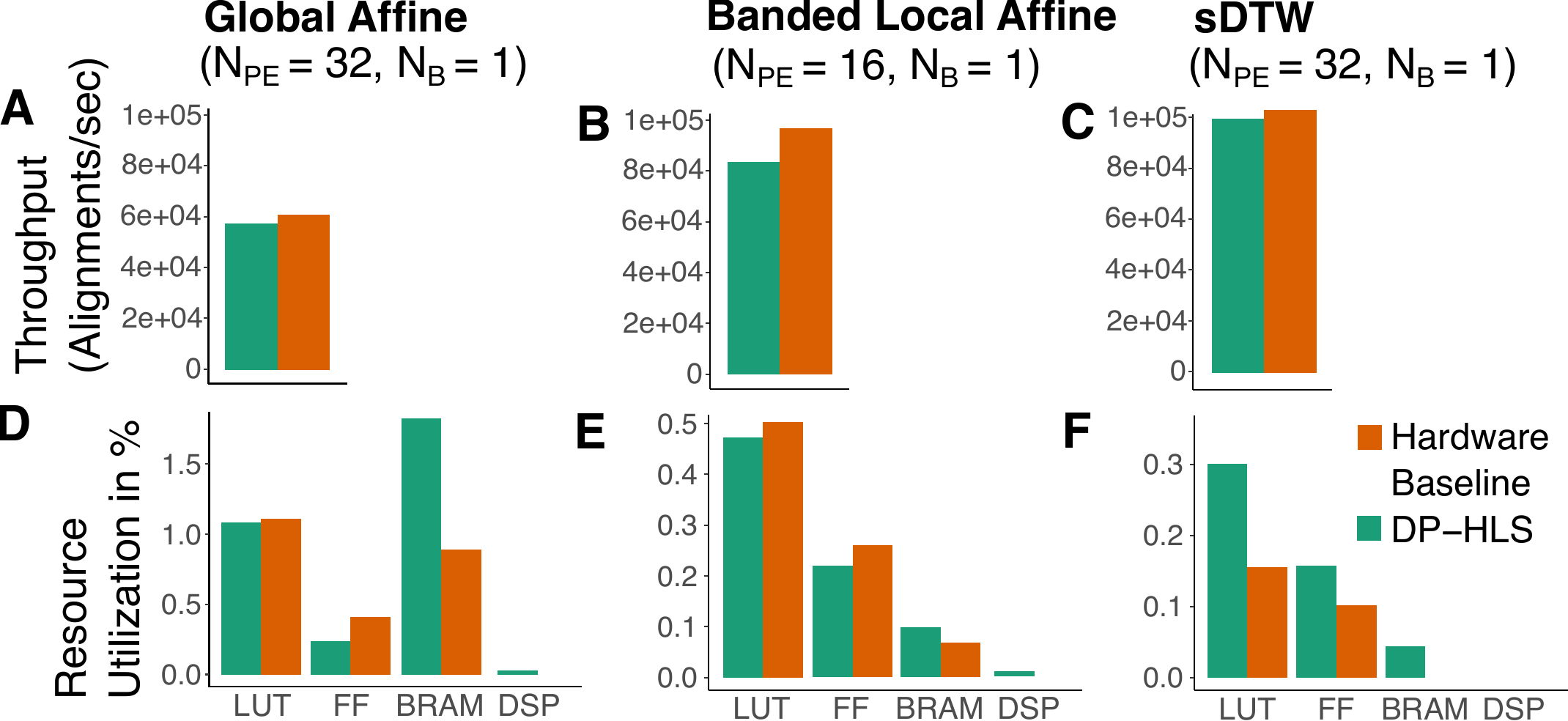}
    \caption{\textbf{Comparison of DP-HLS kernels (\hyperref[kernel:2]{\#2}, \hyperref[kernel:12]{\#12}, \hyperref[kernel:14]{\#14}) with hardware baselines}. (\textbf{A-C}) Throughput, and (\textbf{D-F}) Resource utilization comparison of Kernel \hyperref[kernel:2]{\#2} with GACT \cite{turakhia2018darwin}, Kernel \hyperref[kernel:12]{\#12} with BSW \cite{turakhia2019darwin}, Kernel \hyperref[kernel:14]{\#14} with SquiggleFilter \cite{dunn2021squigglefilter}, respectively.}
    \label{fig:hardware_baseline}
\end{figure}

\begin{figure}
    \includegraphics[width=0.48\textwidth]{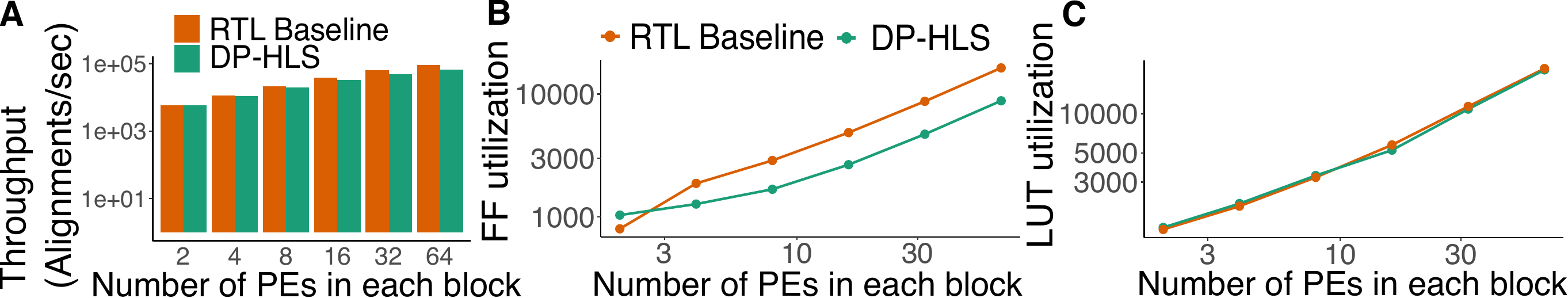}
    \caption{\textbf{Scaling comparison of DP-HLS kernels with hardware baselines (with increasing \texttt{N\textsubscript{PE}, N\textsubscript{B}=1})}. (\textbf{A}) Throughput comparison (in \texttt{log-log} scale), and (\textbf{B-C}) FF and LUT scaling of Global Affine kernel (\hyperref[kernel:2]{\#2}) with its hardware baseline, GACT, respectively.}
    \label{fig:hardware_baseline_scaling}
\end{figure}

Fig. \ref{fig:hardware_baseline}A-C compares the throughput of the three DP-HLS kernels: Global Affine (\hyperref[kernel:2]{\#2}), Banded Local Affine (\hyperref[kernel:12]{\#12}), and sDTW (\hyperref[kernel:14]{\#14}), with the open-source RTL implementations of domain experts, namely GACT~\cite{turakhia2018darwin}, BSW~\cite{turakhia2019darwin} and SquiggleFilter~\cite{dunn2021squigglefilter}. As mentioned in Section~\ref{sec:methods_hardware}, we adjusted the \texttt{N\textsubscript{PE}} and \texttt{N\textsubscript{B}} values in DP-HLS to match those used in the baseline implementations. In all cases, we observed that DP-HLS achieves lower but competitive throughput to hand-crafted RTL implementations. Specifically, the DP-HLS throughput was within 7.7$\%$, 16.8$\%$, and 8.16$\%$ of the baseline for Global Affine, Banded Local Affine, and sDTW kernels, respectively. This is not surprising because, with added flexibility and programming ease, the DP-HLS framework misses out on some optimization opportunities that RTL implementations leverage. For example, all RTL implementations overlap query reads and DP matrix initialization with computation, but these steps are currently performed sequentially in DP-HLS. This overhead is even more apparent in the Banded Local Affine kernel, as it does not employ traceback. While we could perform a similar optimization in DP-HLS, we chose not to implement it, as the benefits are minimal, and it significantly complicates the front-end for end users. The relative throughput of the Global Affine kernel versus GACT remained consistent for long alignments, as both approaches use the same number of tiles. 

The resource utilization comparison is more nuanced. For the Global Affine kernel, DP-HLS shows comparable LUT and FF usage compared to GACT (Fig. \ref{fig:hardware_baseline}D). This is also reflected in the scaling behavior, where throughput remains similar (Fig. \ref{fig:hardware_baseline_scaling}A), and the resource usage difference stays constant (Fig. \ref{fig:hardware_baseline_scaling}B-C). DP-HLS also uses DSPs for pre-computing traceback starting addresses, unlike the baselines, but this does not affect scaling with \texttt{N\textsubscript{PE}}. Despite this, DP-HLS has slightly better LUT and FF utilization on the Banded Local Affine kernel (Fig. \ref{fig:hardware_baseline}E).

Overall, DP-HLS kernels demonstrate efficient resource scaling and competitive throughput compared to optimized RTL designs, with acceptable trade-offs for each kernel.

\subsection{DP-HLS kernels provide 1.3$\times$ - 32$\times$ higher throughput than CPU and GPU baselines}\label{sec:software_baseline_result}

\begin{figure}
    \includegraphics[width=0.40\textwidth]{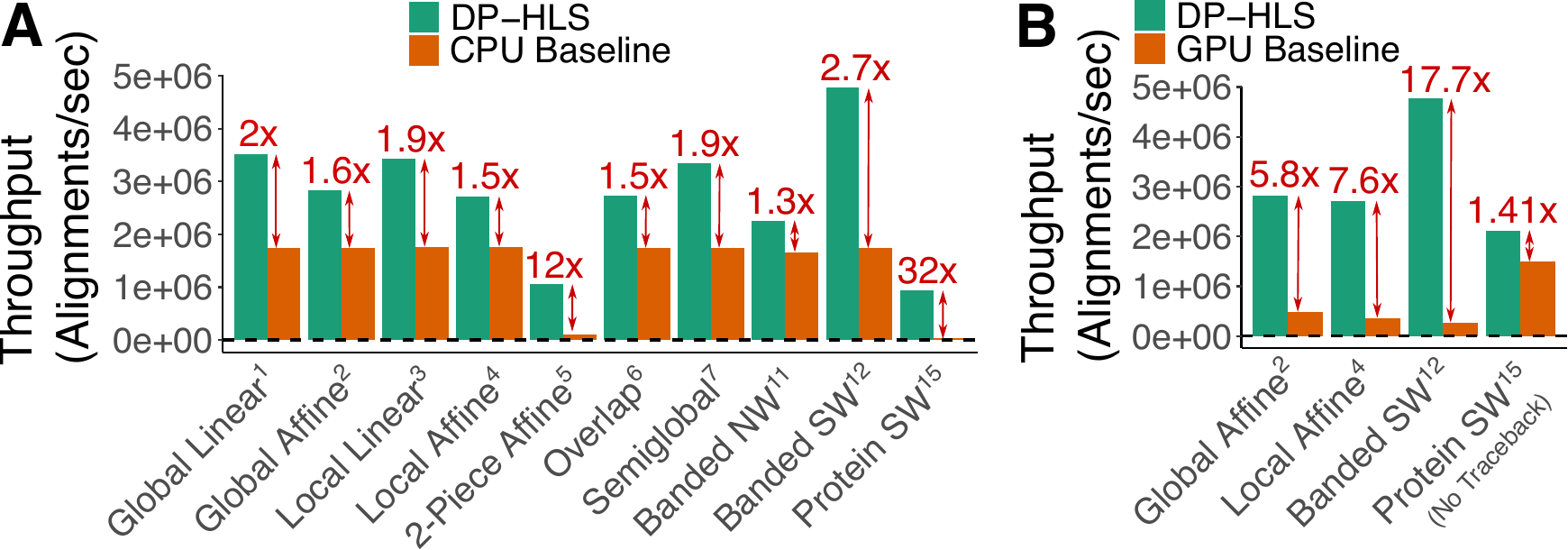}
    \caption{\textbf{Comparison of throughput values of various DP-HLS kernels with their (A) CPU and (B) GPU baselines}. Kernels are referred from Table \ref{table:summary}.}
    \label{fig:software_baseline}
\end{figure}

We performed the iso-cost throughput comparison of DP-HLS with optimized open-source software libraries for applicable kernels on CPU and GPU instances on AWS\textsuperscript{\textregistered} (Section \ref{sec:methods_hardware}). Fig. \ref{fig:software_baseline} summarizes these results. For Kernels \hyperref[kernel:1]{\#1}-\hyperref[kernel:4]{4}, \hyperref[kernel:6]{\#6}-\hyperref[kernel:7]{7}, \hyperref[kernel:11]{\#11}-\hyperref[kernel:12]{12} (Table \ref{table:summary}), the CPU baseline is SeqAn3, which DP-HLS outperforms by 1.5$\times$ to 2.7$\times$ (Fig. \ref{fig:software_baseline}). Interestingly, the baseline throughput shows minor variability across these kernels as SeqAn3 uses the same underlying software implementation across kernels with minor adjustments. In contrast, DP-HLS applies various architectural optimizations tailored to each kernel, maximizing throughput by fitting multiple parallel kernels on the FPGA. For Kernels \hyperref[kernel:5]{\#5} and \hyperref[kernel:15]{\#15}, which are computationally intensive, DP-HLS achieves higher relative throughput, 12$\times$, and 32$\times$, respectively (Fig. \ref{fig:software_baseline}). This boost is likely due to DP-HLS's use of arbitrary precision values and customized hardware data paths, enabling higher throughput than general CPU-based implementations.


Surprisingly, the relative performance of DP-HLS was better even when compared to GPU baselines–1.41$\times$ higher than CUDASW++ and 5.83-17.72$\times$ higher than GASAL2. Since GASAL2 codebase has not been updated recently, it might have been outpaced by newer CPU implementations, particularly in SeqAn3. To a lesser extent, DP-HLS also outperforms CUDASW++, which utilizes modern GPU optimizations, indicating that FPGAs may offer superior efficiency for DP algorithms. This observation is consistent with previous studies \cite{dally_domain-specific_2020}.

\subsection{DP-HLS kernel outperforms a previous HLS baseline}\label{sec:hls_baseline_resilt}

We compared the HLS implementation of the Smith-Waterman algorithm provided in the AMD\textsuperscript{\textregistered} Vitis Genomics Library (\texttt{v2021.2}) with Kernel \hyperref[kernel:3]{\#3} in DP-HLS. DP-HLS achieved 32.6\% higher throughput than the HLS baseline. This difference could be explained by two factors. \emph{First}, the HLS baseline uses streaming functions to transfer some data between the host and device for which the DP-HLS kernels use device memory. \emph{Second}, DP-HLS backend adds more extensive optimization hints to the compiler than the baseline, which results in slightly higher resource utilization than the baseline but better throughput. We also note that it is significantly more challenging for new users to modify the HLS baseline to implement new 2-D DP kernels, as the HLS compiler directives (\texttt{pragma}s) are interleaved with the code that needs to be changed, resulting in a steeper learning curve compared to DP-HLS.

\subsection{DP-HLS can improve the productivity of implementing new DP kernels by \textasciitilde10$\times$}\label{sec:productivity_result}

In addition to providing efficient FPGA implementations for 15 bioinformatically relevant DP kernels, most of which lack existing implementations, the DP-HLS framework aims to simplify the implementation and deployment of new DP kernels.
Based on our experience, setting up the initial DP-HLS framework took several months, as we had to work through back-end optimizations and the quirks of HLS tools. This is similar to the timeline we have experienced for implementing such kernels in HDLs like SystemVerilog. Our experience suggests that current HLS does not necessarily offer a substantial productivity boost over RTL design. However, once the DP-HLS framework was in place, we were able to implement, test, and deploy new 2-D DP kernels on AWS\textsuperscript{\textregistered} EC2 F1 instance in just 2–4 days. Hence, we think that HLS frameworks like DP-HLS, which separate high-level algorithmic specifications in the front-end from low-level optimizations in the back-end, for a broad algorithmic class, such as 2-D DP, can significantly enhance design productivity. For example, a similar framework could be developed for AI applications, mapping diverse matrix multiplication and convolution operations to 2-D systolic arrays~\cite{jouppi2017datacenter, genc2019gemmini, wei2017automated}.

\section{Related Work}
\label{sec:related-work}

\subsection{Dynamic Programming in Bioinformatics}


Dynamic programming (DP) is an algorithmic paradigm that was first introduced in the 1950s by Bellman \cite{bellman1966dynamic}. Needleman-Wunsch \cite{needleman1970general} is one of the earliest and best-known algorithms to have been applied to biological sequence comparison for global pairwise alignment. Since then, the 2-D DP paradigm has become immensely popular in bioinformatics, finding applications in local pairwise alignments \cite{smith1981identification, ukkonen1985finding}, multiple sequence alignment \cite{clustalomega, edgar2004muscle}, homology searches \cite{eddy1998profile, altschul1990basic}, whole-genome alignments \cite{harris2007improved}, basecalling \cite{simpson2017detecting}, variant calling \cite{nielsen2012snp}, and emerging applications like pangenomics \cite{noll2023pangraph, hickey2020genotyping}. Given its computationally intensive nature and broad applicability, NVIDIA\textsuperscript{\textregistered} recently introduced DPX instructions in Hopper GPUs to accelerate the DP paradigm~\cite{elster2022nvidia}. DP-HLS also aims to modularize common 2-D DP algorithmic patterns in a flexible HLS framework, allowing users to easily customize kernels on FPGAs for various bioinformatics applications.

\subsection{Systolic Array Accelerators for Bioinformatics Applications}


Systolic arrays, introduced in the 1970s \cite{kung1979systolic, brent1984systolic}, represent a major advancement in parallel computing by effectively distributing computational tasks across parallel units, with regular patterns of data movement between them. This architecture was initially applied for matrix multiplication \cite{gentleman1982matrix}, and by 1980s, it also found applications in bioinformatics using linear systolic arrays~\cite{lipton1985systolic}. 


In recent years, many linear systolic array based FPGA and ASIC accelerators have been proposed for different bioinformatics applications \cite{fei2018fpgasw, turakhia2018darwin, turakhia2019darwin, haghi2021fpga, zhang2007implementation, msa_dp, bsw_systolic, seedex, dp_2_piece, dp_protein_affine, bwamem_accel, msa_fpga, huang2017hardware, walia2024talco, cali_genasm_2020, cali2022segram}. Examples of such accelerators include GenASM, which accelerates the approximate string matching problem using the Bitap algorithm \cite{cali_genasm_2020}, SquiggleFilter, which accelerates a genomic surveillance application~\cite{dunn2021squigglefilter}, and Darwin-WGA, which accelerates the X-Drop algorithm for whole-genome alignments~\cite{turakhia2019darwin}. Although these accelerators provide huge speedups for specific steps in genomic data analysis, they typically represent a single design point in the DP algorithmic space and cannot be easily refactored into other DP kernels due to their custom low-level HDL implementation. The goal of the DP-HLS framework is to boost design productivity in the development of 2-D DP kernels using HLS while providing competitive performance and similar resources to hand-crafted HDL implementations. 

A recent work, GenDP \cite{gu2023gendp}, shares many similarities with the DP-HLS framework, as it also recognizes that many bioinformatics tasks rely on DP algorithms that are well-suited for systolic array implementations and provides a programmable framework to accelerate these tasks. However, GenDP's programmability operates at the software level, where its processing elements have an instruction set that allows high-level DP algorithm specifications in the form of dataflow-graphs to be compiled. This makes GenDP a software-programmable ASIC. In contrast, DP-HLS targets dataflow architectures, such as FPGAs, where reprogramming involves reconfiguring the hardware itself. Unlike GenDP, DP-HLS kernels can be readily and efficiently deployed on cloud platforms with FPGA support, like AWS\textsuperscript{\textregistered}. We also note that commercial bioinformatics accelerators, such as the DRAGEN\textsuperscript{\texttrademark} Bio-IT Processor by Illumina\textsuperscript{\textregistered}~\cite{goyal2017ultra}, are based on FPGAs, where DP-HLS could be used for providing greater flexibility.

\subsection{The Emergence of HLS Tools and Frameworks}

High-Level Synthesis (HLS) allows users to design custom hardware using high-level languages, significantly reducing hardware development time. As a result, HLS tools and libraries have been widely adopted across various domains \cite{cong_fpga_2022, rupnowStudyHighlevelSynthesis2011}, including machine learning \cite{fahim_hls4ml_2021, shahshahaniFrameworkModelingOptimizing2020}, cryptography \cite{barenghiOpenCLHLSBased2018,homsirikamolNewHLSbasedMethodology2017}, and bioinformatics \cite{hls_read_mapping, hls_chaining, hls_genome_assembly, benkrid_highly_2009}. However, most previous works focus on implementing a single or a handful of kernels for specific applications, offering limited flexibility for users to configure new kernels. An exception in bioinformatics is the work of Benkrid et al. \cite{benkrid_highly_2009}, who developed a flexible systolic array implementation of pairwise sequence alignment algorithms using Handel-C, a high-level HDL. This allowed users to customize hardware designs but only for a limited set of alignment algorithms (global, local, and overlap). While we could not directly compare to Benkrid et al. \cite{benkrid_highly_2009} as their codebase is not actively maintained, DP-HLS offers much greater flexibility within the 2-D DP paradigm, allowing users to also configure scoring functions, scoring layers, input characters, traceback logic, and more.

\section{Conclusion and Future Work}
\label{sec:conclusion}

In this paper, we introduce DP-HLS, a novel HLS-based framework designed for FPGA acceleration of 2-D DP algorithms that are widely used in bioinformatics applications. DP-HLS allows users without hardware design experience to quickly develop efficient hardware solutions for DP algorithms with high flexibility. DP-HLS does this by introducing an abstraction layer in HLS design, which takes advantage of the fact that a broad range of applications can efficiently map to the same hardware primitive. DP-HLS is deployed on the AWS\textsuperscript{\textregistered} cloud platform and achieves competitive throughput and similar resource utilization to hand-crafted RTL. In the future, we plan to implement HLS frameworks for other algorithmic classes, such as those used in AI and signal processing, using similar strategies.


\section{Acknowledgments}
We thank the AMD-omics group for helpful feedback. Research reported in this manuscript was supported by an Amazon Research Award (Fall 2022 CFP), AMD AI \& HPC Fund, and the Hellman Fellowship.

\bibliographystyle{IEEEtran}
\bibliography{references}

\begin{thebibliography}{10}
\providecommand{\url}[1]{#1}
\csname url@samestyle\endcsname
\providecommand{\newblock}{\relax}
\providecommand{\bibinfo}[2]{#2}
\providecommand{\BIBentrySTDinterwordspacing}{\spaceskip=0pt\relax}
\providecommand{\BIBentryALTinterwordstretchfactor}{4}
\providecommand{\BIBentryALTinterwordspacing}{\spaceskip=\fontdimen2\font plus
\BIBentryALTinterwordstretchfactor\fontdimen3\font minus
  \fontdimen4\font\relax}
\providecommand{\BIBforeignlanguage}[2]{{%
\expandafter\ifx\csname l@#1\endcsname\relax
\typeout{** WARNING: IEEEtran.bst: No hyphenation pattern has been}%
\typeout{** loaded for the language `#1'. Using the pattern for}%
\typeout{** the default language instead.}%
\else
\language=\csname l@#1\endcsname
\fi
#2}}
\providecommand{\BIBdecl}{\relax}
\BIBdecl

\bibitem{stephens2015big}
Z.~D. Stephens, S.~Y. Lee, F.~Faghri, R.~H. Campbell, C.~Zhai, M.~J. Efron,
  R.~Iyer, M.~C. Schatz, S.~Sinha, and G.~E. Robinson, ``Big data: astronomical
  or genomical?'' \emph{PLoS biology}, vol.~13, no.~7, p. e1002195, 2015.

\bibitem{lindegger2023scrooge}
J.~Lindegger, D.~Senol~Cali, M.~Alser, J.~G{\'o}mez-Luna, N.~M. Ghiasi, and
  O.~Mutlu, ``Scrooge: a fast and memory-frugal genomic sequence aligner for
  cpus, gpus, and asics,'' \emph{Bioinformatics}, vol.~39, no.~5, p. btad151,
  2023.

\bibitem{ahmed2020gpu}
N.~Ahmed, T.~D. Qiu, K.~Bertels, and Z.~Al-Ars, ``Gpu acceleration of darwin
  read overlapper for de novo assembly of long dna reads,'' \emph{BMC
  bioinformatics}, vol.~21, pp. 1--17, 2020.

\bibitem{zeni2020logan}
A.~Zeni, G.~Guidi, M.~Ellis, N.~Ding, M.~D. Santambrogio, S.~Hofmeyr,
  A.~Bulu{\c{c}}, L.~Oliker, and K.~Yelick, ``Logan: High-performance gpu-based
  x-drop long-read alignment,'' in \emph{2020 IEEE International Parallel and
  Distributed Processing Symposium (IPDPS)}.\hskip 1em plus 0.5em minus
  0.4em\relax IEEE, 2020, pp. 462--471.

\bibitem{ahmed2019gasal2}
N.~Ahmed, J.~L{\'e}vy, S.~Ren, H.~Mushtaq, K.~Bertels, and Z.~Al-Ars, ``Gasal2:
  a gpu accelerated sequence alignment library for high-throughput ngs data,''
  \emph{BMC bioinformatics}, vol.~20, pp. 1--20, 2019.

\bibitem{goenka2020segalign}
S.~D. Goenka, Y.~Turakhia, B.~Paten, and M.~Horowitz, ``Segalign: A scalable
  gpu-based whole genome aligner,'' in \emph{SC20: International Conference for
  High Performance Computing, Networking, Storage and Analysis}.\hskip 1em plus
  0.5em minus 0.4em\relax IEEE, 2020, pp. 1--13.

\bibitem{park2024agatha}
S.~Park, J.~Hong, J.~Song, H.~Kim, Y.~Kim, and J.~Lee, ``Agatha: Fast and
  efficient gpu acceleration of guided sequence alignment for long read
  mapping,'' in \emph{Proceedings of the 29th ACM SIGPLAN Annual Symposium on
  Principles and Practice of Parallel Programming}, 2024, pp. 431--444.

\bibitem{aguado2022wfa}
Q.~Aguado-Puig, S.~Marco-Sola, J.~C. Moure, C.~Matzoros, D.~Castells-Rufas,
  A.~Espinosa, and M.~Moreto, ``Wfa-gpu: Gap-affine pairwise alignment using
  gpus,'' \emph{bioRxiv}, pp. 2022--04, 2022.

\bibitem{zenigpu2024}
\BIBentryALTinterwordspacing
A.~Zeni, S.~Onken, M.~D. Santambrogio, and M.~Samadi, ``Leveraging difference
  recurrence relations for high-performance gpu genome alignment,'' in
  \emph{Proceedings of the 2024 International Conference on Parallel
  Architectures and Compilation Techniques}, ser. PACT '24.\hskip 1em plus
  0.5em minus 0.4em\relax New York, NY, USA: Association for Computing
  Machinery, 2024, p. 133–143. [Online]. Available:
  \url{https://doi.org/10.1145/3656019.3676894}
\BIBentrySTDinterwordspacing

\bibitem{fei2018fpgasw}
X.~Fei, Z.~Dan, L.~Lina, M.~Xin, and Z.~Chunlei, ``Fpgasw: accelerating
  large-scale smith--waterman sequence alignment application with backtracking
  on fpga linear systolic array,'' \emph{Interdisciplinary Sciences:
  Computational Life Sciences}, vol.~10, pp. 176--188, 2018.

\bibitem{turakhia2018darwin}
Y.~Turakhia, G.~Bejerano, and W.~J. Dally, ``Darwin: A genomics co-processor
  provides up to 15,000 x acceleration on long read assembly,'' \emph{ACM
  SIGPLAN Notices}, vol.~53, no.~2, pp. 199--213, 2018.

\bibitem{turakhia2019darwin}
Y.~Turakhia, S.~D. Goenka, G.~Bejerano, and W.~J. Dally, ``Darwin-wga: A
  co-processor provides increased sensitivity in whole genome alignments with
  high speedup,'' in \emph{2019 IEEE International Symposium on High
  Performance Computer Architecture (HPCA)}.\hskip 1em plus 0.5em minus
  0.4em\relax IEEE, 2019, pp. 359--372.

\bibitem{haghi2021fpga}
A.~Haghi, S.~Marco-Sola, L.~Alvarez, D.~Diamantopoulos, C.~Hagleitner, and
  M.~Moreto, ``An fpga accelerator of the wavefront algorithm for genomics
  pairwise alignment,'' in \emph{2021 31st International Conference on
  Field-Programmable Logic and Applications (FPL)}.\hskip 1em plus 0.5em minus
  0.4em\relax IEEE, 2021, pp. 151--159.

\bibitem{zhang2007implementation}
P.~Zhang, G.~Tan, and G.~R. Gao, ``Implementation of the smith-waterman
  algorithm on a reconfigurable supercomputing platform,'' in \emph{Proceedings
  of the 1st international workshop on High-performance reconfigurable
  computing technology and applications: held in conjunction with SC07}, 2007,
  pp. 39--48.

\bibitem{msa_dp}
R.-T. Chien, Y.-L. Liao, C.-A. Wang, Y.-C. Li, and Y.-C. Lu,
  ``Three-dimensional dynamic programming accelerator for multiple sequence
  alignment,'' in \emph{2018 IEEE Nordic Circuits and Systems Conference
  (NORCAS): NORCHIP and International Symposium of System-on-Chip (SoC)}, 2018,
  pp. 1--5.

\bibitem{bsw_systolic}
P.~Chen, C.~Wang, X.~Li, and X.~Zhou, ``Hardware acceleration for the banded
  smith-waterman algorithm with the cycled systolic array,'' in \emph{2013
  International Conference on Field-Programmable Technology (FPT)}, 2013, pp.
  480--481.

\bibitem{seedex}
D.~Fujiki, S.~Wu, N.~Ozog, K.~Goliya, D.~Blaauw, S.~Narayanasamy, and R.~Das,
  ``Seedex: A genome sequencing accelerator for optimal alignments in
  subminimal space,'' in \emph{2020 53rd Annual IEEE/ACM International
  Symposium on Microarchitecture (MICRO)}, 2020, pp. 937--950.

\bibitem{dp_2_piece}
J.-P. Wu, Y.-C. Lin, Y.-W. Wu, S.-W. Hsieh, C.-H. Tai, and Y.-C. Lu, ``A
  memory-efficient accelerator for dna sequence alignment with two-piece affine
  gap tracebacks,'' in \emph{2021 IEEE International Symposium on Circuits and
  Systems (ISCAS)}, 2021, pp. 1--4.

\bibitem{dp_protein_affine}
M.-J. Lin, Y.-C. Li, and Y.-C. Lu, ``Hardware accelerator design for
  dynamic-programming-based protein sequence alignment with affine gap
  tracebacks,'' in \emph{2019 IEEE Biomedical Circuits and Systems Conference
  (BioCAS)}, 2019, pp. 1--4.

\bibitem{bwamem_accel}
E.~J. Houtgast, V.-M. Sima, K.~Bertels, and Z.~Al-Ars, ``An fpga-based systolic
  array to accelerate the bwa-mem genomic mapping algorithm,'' in \emph{2015
  International Conference on Embedded Computer Systems: Architectures,
  Modeling, and Simulation (SAMOS)}, 2015, pp. 221--227.

\bibitem{msa_fpga}
T.~Oliver, B.~Schmidt, D.~Maskell, D.~Nathan, and R.~Clemens, ``Multiple
  sequence alignment on an fpga,'' in \emph{11th International Conference on
  Parallel and Distributed Systems (ICPADS'05)}, vol.~2, 2005, pp. 326--330.

\bibitem{huang2017hardware}
S.~Huang, G.~J. Manikandan, A.~Ramachandran, K.~Rupnow, W.-m.~W. Hwu, and
  D.~Chen, ``Hardware acceleration of the pair-hmm algorithm for dna variant
  calling,'' in \emph{Proceedings of the 2017 ACM/SIGDA International Symposium
  on Field-Programmable Gate Arrays}, 2017, pp. 275--284.

\bibitem{walia2024talco}
S.~Walia, C.~Ye, A.~Bera, D.~Lodhavia, and Y.~Turakhia, ``Talco: Tiling genome
  sequence alignment using convergence of traceback pointers,'' in \emph{2024
  IEEE International Symposium on High-Performance Computer Architecture
  (HPCA)}.\hskip 1em plus 0.5em minus 0.4em\relax IEEE, 2024, pp. 91--107.

\bibitem{cali_genasm_2020}
\BIBentryALTinterwordspacing
D.~S. Cali, G.~S. Kalsi, Z.~Bingöl, C.~Firtina, L.~Subramanian, J.~S. Kim,
  R.~Ausavarungnirun, M.~Alser, J.~Gomez-Luna, A.~Boroumand, A.~Nori,
  A.~Scibisz, S.~Subramoney, C.~Alkan, S.~Ghose, and O.~Mutlu,
  ``\BIBforeignlanguage{en}{{GenASM}: {A} {High}-{Performance}, {Low}-{Power}
  {Approximate} {String} {Matching} {Acceleration} {Framework} for {Genome}
  {Sequence} {Analysis}},'' Sep. 2020, arXiv:2009.07692 [cs, q-bio]. [Online].
  Available: \url{http://arxiv.org/abs/2009.07692}
\BIBentrySTDinterwordspacing

\bibitem{cali2022segram}
D.~S. Cali, K.~Kanellopoulos, J.~Lindegger, Z.~Bing{\"o}l, G.~S. Kalsi, Z.~Zuo,
  C.~Firtina, M.~B. Cavlak, J.~Kim, N.~M. Ghiasi \emph{et~al.}, ``Segram: A
  universal hardware accelerator for genomic sequence-to-graph and
  sequence-to-sequence mapping,'' in \emph{Proceedings of the 49th Annual
  International Symposium on Computer Architecture}, 2022, pp. 638--655.

\bibitem{bellman1966dynamic}
R.~Bellman, ``Dynamic programming,'' \emph{science}, vol. 153, no. 3731, pp.
  34--37, 1966.

\bibitem{smith1981identification}
T.~F. Smith, M.~S. Waterman \emph{et~al.}, ``Identification of common molecular
  subsequences,'' \emph{Journal of molecular biology}, vol. 147, no.~1, pp.
  195--197, 1981.

\bibitem{ukkonen1985finding}
E.~Ukkonen, ``Finding approximate patterns in strings,'' \emph{Journal of
  algorithms}, vol.~6, no.~1, pp. 132--137, 1985.

\bibitem{clustalomega}
F.~Sievers, A.~Wilm, D.~Dineen, T.~J. Gibson, K.~Karplus, W.~Li, R.~Lopez,
  H.~McWilliam, M.~Remmert, J.~S{\"o}ding \emph{et~al.}, ``Fast, scalable
  generation of high-quality protein multiple sequence alignments using clustal
  omega,'' \emph{Molecular systems biology}, vol.~7, no.~1, p. 539, 2011.

\bibitem{edgar2004muscle}
R.~C. Edgar, ``Muscle: a multiple sequence alignment method with reduced time
  and space complexity,'' \emph{BMC bioinformatics}, vol.~5, pp. 1--19, 2004.

\bibitem{eddy1998profile}
S.~R. Eddy, ``Profile hidden markov models.'' \emph{Bioinformatics (Oxford,
  England)}, vol.~14, no.~9, pp. 755--763, 1998.

\bibitem{altschul1990basic}
S.~F. Altschul, W.~Gish, W.~Miller, E.~W. Myers, and D.~J. Lipman, ``Basic
  local alignment search tool,'' \emph{Journal of molecular biology}, vol. 215,
  no.~3, pp. 403--410, 1990.

\bibitem{harris2007improved}
R.~S. Harris, \emph{Improved pairwise alignment of genomic DNA}.\hskip 1em plus
  0.5em minus 0.4em\relax The Pennsylvania State University, 2007.

\bibitem{simpson2017detecting}
J.~T. Simpson, R.~E. Workman, P.~Zuzarte, M.~David, L.~Dursi, and W.~Timp,
  ``Detecting dna cytosine methylation using nanopore sequencing,''
  \emph{Nature methods}, vol.~14, no.~4, pp. 407--410, 2017.

\bibitem{nielsen2012snp}
R.~Nielsen, T.~Korneliussen, A.~Albrechtsen, Y.~Li, and J.~Wang, ``Snp calling,
  genotype calling, and sample allele frequency estimation from new-generation
  sequencing data,'' 2012.

\bibitem{subramaniyan2021genomicsbench}
A.~Subramaniyan, Y.~Gu, T.~Dunn, S.~Paul, M.~Vasimuddin, S.~Misra, D.~Blaauw,
  S.~Narayanasamy, and R.~Das, ``Genomicsbench: A benchmark suite for
  genomics,'' in \emph{2021 IEEE International Symposium on Performance
  Analysis of Systems and Software (ISPASS)}.\hskip 1em plus 0.5em minus
  0.4em\relax IEEE, 2021, pp. 1--12.

\bibitem{elster2022nvidia}
A.~C. Elster and T.~A. Haugdahl, ``Nvidia hopper gpu and grace cpu
  highlights,'' \emph{Computing in Science \& Engineering}, vol.~24, no.~2, pp.
  95--100, 2022.

\bibitem{lipton1985systolic}
R.~J. Lipton and D.~Lopresti, ``A systolic array for rapid string comparison,''
  in \emph{Proceedings of the Chapel Hill conference on VLSI}.\hskip 1em plus
  0.5em minus 0.4em\relax Chapel Hill NC, 1985, pp. 363--376.

\bibitem{gu2023gendp}
Y.~Gu, A.~Subramaniyan, T.~Dunn, A.~Khadem, K.-Y. Chen, S.~Paul, M.~Vasimuddin,
  S.~Misra, D.~Blaauw, S.~Narayanasamy \emph{et~al.}, ``Gendp: A framework of
  dynamic programming acceleration for genome sequencing analysis,'' in
  \emph{Proceedings of the 50th Annual International Symposium on Computer
  Architecture}, 2023, pp. 1--15.

\bibitem{behera2024comprehensive}
S.~Behera, S.~Catreux, M.~Rossi, S.~Truong, Z.~Huang, M.~Ruehle, A.~Visvanath,
  G.~Parnaby, C.~Roddey, V.~Onuchic \emph{et~al.}, ``Comprehensive genome
  analysis and variant detection at scale using dragen,'' \emph{Nature
  Biotechnology}, pp. 1--15, 2024.

\bibitem{goyal2017ultra}
A.~Goyal, H.~J. Kwon, K.~Lee, R.~Garg, S.~Y. Yun, Y.~H. Kim, S.~Lee, and M.~S.
  Lee, ``Ultra-fast next generation human genome sequencing data processing
  using dragentm bio-it processor for precision medicine,'' \emph{Open Journal
  of Genetics}, vol.~7, no.~1, pp. 9--19, 2017.

\bibitem{timelogic}
\BIBentryALTinterwordspacing
{TimeLogic}® biocomputing solutions. [Online]. Available:
  \url{https://www.activemotif.com/catalog/75/timelogic-biocomputing-solutions}
\BIBentrySTDinterwordspacing

\bibitem{ncbi_blast}
{NCBI}, ``Blast: Basic local alignment search tool,''
  \url{https://blast.ncbi.nlm.nih.gov/Blast.cgi}, 2024, accessed: 2024-07-23.

\bibitem{PMID:38597606}
\BIBentryALTinterwordspacing
F.~Madeira, N.~Madhusoodanan, J.~Lee, A.~Eusebi, A.~Niewielska, A.~R.~N. Tivey,
  R.~Lopez, and S.~Butcher, ``The embl-ebi job dispatcher sequence analysis
  tools framework in 2024,'' \emph{Nucleic acids research}, p. gkae241, April
  2024. [Online]. Available:
  \url{https://academic.oup.com/nar/advance-article-pdf/doi/10.1093/nar/gkae241/57199791/gkae241.pdf}
\BIBentrySTDinterwordspacing

\bibitem{fasta}
W.~R. Pearson, ``Using the fasta program to search protein and dna sequence
  databases,'' \emph{Computer Analysis of Sequence Data: Part I}, pp. 307--331,
  1994.

\bibitem{kent2002blat}
W.~J. Kent, ``Blat—the blast-like alignment tool,'' \emph{Genome research},
  vol.~12, no.~4, pp. 656--664, 2002.

\bibitem{lastz}
R.~S. Harris, \emph{Improved pairwise alignment of genomic DNA}.\hskip 1em plus
  0.5em minus 0.4em\relax The Pennsylvania State University, 2007.

\bibitem{li_minimap2_2018}
H.~Li, ``Minimap2: {Pairwise} alignment for nucleotide sequences,''
  \emph{Bioinformatics}, vol.~34, no.~18, pp. 3094--3100, Sep. 2018, arXiv:
  1708.01492 Publisher: Oxford University Press.

\bibitem{canu}
S.~Koren, B.~P. Walenz, K.~Berlin, J.~R. Miller, N.~H. Bergman, and A.~M.
  Phillippy, ``Canu: scalable and accurate long-read assembly via adaptive
  k-mer weighting and repeat separation,'' \emph{Genome research}, vol.~27,
  no.~5, pp. 722--736, 2017.

\bibitem{flye}
M.~Kolmogorov, J.~Yuan, Y.~Lin, and P.~A. Pevzner, ``Assembly of long,
  error-prone reads using repeat graphs,'' \emph{Nature biotechnology},
  vol.~37, no.~5, pp. 540--546, 2019.

\bibitem{bwa-mem}
H.~Li, ``Aligning sequence reads, clone sequences and assembly contigs with
  bwa-mem,'' \emph{arXiv preprint arXiv:1303.3997}, 2013.

\bibitem{thompsonCLUSTALImprovingSensitivity1994}
J.~D. Thompson, D.~G. Higgins, and T.~J. Gibson, ``{{CLUSTAL W}}: Improving the
  sensitivity of progressive multiple sequence alignment through sequence
  weighting, position-specific gap penalties and weight matrix choice.''
  \emph{Nucleic Acids Research}, vol.~22, no.~22, pp. 4673--4680, Nov. 1994.

\bibitem{squigglekit}
\BIBentryALTinterwordspacing
J.~M. Ferguson and M.~A. Smith, ``{SquiggleKit: a toolkit for manipulating
  nanopore signal data},'' \emph{Bioinformatics}, vol.~35, no.~24, pp.
  5372--5373, 07 2019. [Online]. Available:
  \url{https://doi.org/10.1093/bioinformatics/btz586}
\BIBentrySTDinterwordspacing

\bibitem{finnHMMERWebServer2011}
R.~D. Finn, J.~Clements, and S.~R. Eddy, ``{{HMMER}} web server: Interactive
  sequence similarity searching,'' \emph{Nucleic Acids Research}, vol.~39, no.
  Web Server issue, pp. W29--W37, Jul. 2011.

\bibitem{stanke2005augustus}
M.~Stanke and B.~Morgenstern, ``Augustus: a web server for gene prediction in
  eukaryotes that allows user-defined constraints,'' \emph{Nucleic acids
  research}, vol.~33, no. suppl\_2, pp. W465--W467, 2005.

\bibitem{bowtie}
B.~Langmead and S.~L. Salzberg, ``Fast gapped-read alignment with bowtie 2,''
  \emph{Nature methods}, vol.~9, no.~4, pp. 357--359, 2012.

\bibitem{dunn2021squigglefilter}
T.~Dunn, H.~Sadasivan, J.~Wadden, K.~Goliya, K.-Y. Chen, D.~Blaauw, R.~Das, and
  S.~Narayanasamy, ``Squigglefilter: An accelerator for portable virus
  detection,'' in \emph{MICRO-54: 54th Annual IEEE/ACM International Symposium
  on Microarchitecture}, 2021, pp. 535--549.

\bibitem{rawhash}
\BIBentryALTinterwordspacing
C.~Firtina, N.~Mansouri~Ghiasi, J.~Lindegger, G.~Singh, M.~B. Cavlak, H.~Mao,
  and O.~Mutlu, ``Rawhash: enabling fast and accurate real-time analysis of raw
  nanopore signals for large genomes,'' \emph{Bioinformatics}, vol.~39, no.
  Supplement\_1, pp. 297--307, 06 2023. [Online]. Available:
  \url{https://doi.org/10.1093/bioinformatics/btad272}
\BIBentrySTDinterwordspacing

\bibitem{buchfink2021sensitive}
B.~Buchfink, K.~Reuter, and H.-G. Drost, ``Sensitive protein alignments at
  tree-of-life scale using diamond,'' \emph{Nature methods}, vol.~18, no.~4,
  pp. 366--368, 2021.

\bibitem{pal2006evolutionary}
S.~K. Pal, S.~Bandyopadhyay, and S.~S. Ray, ``Evolutionary computation in
  bioinformatics: A review,'' \emph{IEEE Transactions on Systems, Man, and
  Cybernetics, Part C (Applications and Reviews)}, vol.~36, no.~5, pp.
  601--615, 2006.

\bibitem{cohen2004bioinformatics}
J.~Cohen, ``Bioinformatics—an introduction for computer scientists,''
  \emph{ACM Computing Surveys (CSUR)}, vol.~36, no.~2, pp. 122--158, 2004.

\bibitem{needleman1970general}
S.~B. Needleman and C.~D. Wunsch, ``A general method applicable to the search
  for similarities in the amino acid sequence of two proteins,'' \emph{Journal
  of molecular biology}, vol.~48, no.~3, pp. 443--453, 1970.

\bibitem{wangScoringProfiletoprofileSequence2004}
G.~Wang and R.~L. Dunbrack, ``Scoring profile-to-profile sequence alignments,''
  \emph{Protein Science : A Publication of the Protein Society}, vol.~13,
  no.~6, pp. 1612--1626, Jun. 2004.

\bibitem{bellman1959adaptive}
R.~Bellman and R.~Kalaba, ``On adaptive control processes,'' \emph{IRE
  Transactions on Automatic Control}, vol.~4, no.~2, pp. 1--9, 1959.

\bibitem{han2018accurate}
R.~Han, Y.~Li, X.~Gao, and S.~Wang, ``An accurate and rapid continuous wavelet
  dynamic time warping algorithm for end-to-end mapping in ultra-long nanopore
  sequencing,'' \emph{Bioinformatics}, vol.~34, no.~17, pp. i722--i731, 2018.

\bibitem{henikoff_amino_1992}
\BIBentryALTinterwordspacing
S.~Henikoff and J.~G. Henikoff, ``Amino acid substitution matrices from protein
  blocks.'' \emph{Proceedings of the National Academy of Sciences of the United
  States of America}, vol.~89, no.~22, pp. 10\,915--10\,919, Nov. 1992.
  [Online]. Available:
  \url{https://www.ncbi.nlm.nih.gov/pmc/articles/PMC50453/}
\BIBentrySTDinterwordspacing

\bibitem{skutkova_classification_2013}
\BIBentryALTinterwordspacing
H.~Skutkova, M.~Vitek, P.~Babula, R.~Kizek, and I.~Provaznik,
  ``\BIBforeignlanguage{en}{Classification of genomic signals using dynamic
  time warping},'' \emph{\BIBforeignlanguage{en}{BMC Bioinformatics}}, vol.~14,
  no. S10, p.~S1, Aug. 2013. [Online]. Available:
  \url{https://bmcbioinformatics.biomedcentral.com/articles/10.1186/1471-2105-14-S10-S1}
\BIBentrySTDinterwordspacing

\bibitem{gotoh_improved_1982}
\BIBentryALTinterwordspacing
O.~Gotoh, ``An improved algorithm for matching biological sequences,''
  \emph{Journal of Molecular Biology}, vol. 162, no.~3, pp. 705--708, Dec.
  1982. [Online]. Available:
  \url{https://www.sciencedirect.com/science/article/pii/0022283682903989}
\BIBentrySTDinterwordspacing

\bibitem{chaoAligningTwoSequences1992}
K.-M. Chao, W.~R. Pearson, and W.~Miller, ``Aligning two sequences within a
  specified diagonal band,'' \emph{Bioinformatics}, vol.~8, no.~5, pp.
  481--487, Oct. 1992.

\bibitem{zhangGreedyAlgorithmAligning2000}
Z.~Zhang, S.~Schwartz, L.~Wagner, and W.~Miller, ``A {{Greedy Algorithm}} for
  {{Aligning DNA Sequences}},'' \emph{Journal of Computational Biology},
  vol.~7, no. 1-2, pp. 203--214, Feb. 2000.

\bibitem{VitisHighLevelSynthesis2021}
``Vitis {{High-Level Synthesis User Guide}},'' 2021.

\bibitem{munshi2011opencl}
A.~Munshi, B.~Gaster, T.~G. Mattson, and D.~Ginsburg, \emph{OpenCL programming
  guide}.\hskip 1em plus 0.5em minus 0.4em\relax Pearson Education, 2011.

\bibitem{ono_pbsim2_2021}
\BIBentryALTinterwordspacing
Y.~Ono, K.~Asai, and M.~Hamada, ``{PBSIM2}: a simulator for long-read
  sequencers with a novel generative model of quality scores,''
  \emph{Bioinformatics}, vol.~37, no.~5, pp. 589--595, Mar. 2021. [Online].
  Available: \url{https://doi.org/10.1093/bioinformatics/btaa835}
\BIBentrySTDinterwordspacing

\bibitem{rhoads_pacbio_2015}
A.~Rhoads and K.~F. Au, ``\BIBforeignlanguage{eng}{{PacBio} {Sequencing} and
  {Its} {Applications}},'' \emph{\BIBforeignlanguage{eng}{Genomics, Proteomics
  \& Bioinformatics}}, vol.~13, no.~5, pp. 278--289, Oct. 2015.

\bibitem{the_uniprot_consortium_uniprot_2023}
\BIBentryALTinterwordspacing
{The UniProt Consortium}, A.~Bateman, M.-J. Martin, S.~Orchard, M.~Magrane,
  S.~Ahmad, E.~Alpi, E.~H. Bowler-Barnett, R.~Britto, H.~Bye-A-Jee, A.~Cukura,
  P.~Denny, T.~Dogan, T.~Ebenezer, J.~Fan, P.~Garmiri, L.~J. Da~Costa~Gonzales,
  E.~Hatton-Ellis, A.~Hussein, A.~Ignatchenko, G.~Insana, R.~Ishtiaq, V.~Joshi,
  D.~Jyothi, S.~Kandasaamy, A.~Lock, A.~Luciani, M.~Lugaric, J.~Luo, Y.~Lussi,
  A.~MacDougall, F.~Madeira, M.~Mahmoudy, A.~Mishra, K.~Moulang,
  A.~Nightingale, S.~Pundir, G.~Qi, S.~Raj, P.~Raposo, D.~L. Rice, R.~Saidi,
  R.~Santos, E.~Speretta, J.~Stephenson, P.~Totoo, E.~Turner, N.~Tyagi,
  P.~Vasudev, K.~Warner, X.~Watkins, R.~Zaru, H.~Zellner, A.~J. Bridge,
  L.~Aimo, G.~Argoud-Puy, A.~H. Auchincloss, K.~B. Axelsen, P.~Bansal,
  D.~Baratin, T.~M. Batista~Neto, M.-C. Blatter, J.~T. Bolleman, E.~Boutet,
  L.~Breuza, B.~C. Gil, C.~Casals-Casas, K.~C. Echioukh, E.~Coudert, B.~Cuche,
  E.~De~Castro, A.~Estreicher, M.~L. Famiglietti, M.~Feuermann, E.~Gasteiger,
  P.~Gaudet, S.~Gehant, V.~Gerritsen, A.~Gos, N.~Gruaz, C.~Hulo,
  N.~Hyka-Nouspikel, F.~Jungo, A.~Kerhornou, P.~Le~Mercier, D.~Lieberherr,
  P.~Masson, A.~Morgat, V.~Muthukrishnan, S.~Paesano, I.~Pedruzzi, S.~Pilbout,
  L.~Pourcel, S.~Poux, M.~Pozzato, M.~Pruess, N.~Redaschi, C.~Rivoire, C.~J.~A.
  Sigrist, K.~Sonesson, S.~Sundaram, C.~H. Wu, C.~N. Arighi, L.~Arminski,
  C.~Chen, Y.~Chen, H.~Huang, K.~Laiho, P.~McGarvey, D.~A. Natale, K.~Ross,
  C.~R. Vinayaka, Q.~Wang, Y.~Wang, and J.~Zhang,
  ``\BIBforeignlanguage{en}{{UniProt}: the {Universal} {Protein}
  {Knowledgebase} in 2023},'' \emph{\BIBforeignlanguage{en}{Nucleic Acids
  Research}}, vol.~51, no.~D1, pp. D523--D531, Jan. 2023. [Online]. Available:
  \url{https://academic.oup.com/nar/article/51/D1/D523/6835362}
\BIBentrySTDinterwordspacing

\bibitem{reinertSeqAnTemplateLibrary2017}
K.~Reinert, T.~H. Dadi, M.~Ehrhardt, H.~Hauswedell, S.~Mehringer, R.~Rahn,
  J.~Kim, C.~Pockrandt, J.~Winkler, E.~Siragusa, G.~Urgese, and D.~Weese, ``The
  {{SeqAn C}}++ template library for efficient sequence analysis: {{A}}
  resource for programmers,'' \emph{Journal of Biotechnology}, vol. 261, pp.
  157--168, Nov. 2017.

\bibitem{schmidt2023cudasw++}
B.~Schmidt, F.~Kallenborn, A.~Chacon, and C.~Hundt, ``Cudasw++ 4.0: ultra-fast
  gpu-based smith-waterman protein sequence database search,'' \emph{bioRxiv},
  pp. 2023--10, 2023.

\bibitem{xilinx_vitis_libraries}
\BIBentryALTinterwordspacing
Xilinx, ``Xilinx vitis libraries.'' [Online]. Available:
  \url{https://xilinx.github.io/Vitis_Libraries}
\BIBentrySTDinterwordspacing

\bibitem{dally_domain-specific_2020}
\BIBentryALTinterwordspacing
W.~J. Dally, Y.~Turakhia, and S.~Han, ``\BIBforeignlanguage{en}{Domain-specific
  hardware accelerators},'' \emph{\BIBforeignlanguage{en}{Communications of the
  ACM}}, vol.~63, no.~7, pp. 48--57, Jun. 2020. [Online]. Available:
  \url{https://dl.acm.org/doi/10.1145/3361682}
\BIBentrySTDinterwordspacing

\bibitem{jouppi2017datacenter}
N.~P. Jouppi, C.~Young, N.~Patil, D.~Patterson, G.~Agrawal, R.~Bajwa, S.~Bates,
  S.~Bhatia, N.~Boden, A.~Borchers \emph{et~al.}, ``In-datacenter performance
  analysis of a tensor processing unit,'' in \emph{Proceedings of the 44th
  annual international symposium on computer architecture}, 2017, pp. 1--12.

\bibitem{genc2019gemmini}
H.~Genc, A.~Haj-Ali, V.~Iyer, A.~Amid, H.~Mao, J.~Wright, C.~Schmidt, J.~Zhao,
  A.~Ou, M.~Banister \emph{et~al.}, ``Gemmini: An agile systolic array
  generator enabling systematic evaluations of deep-learning architectures,''
  \emph{arXiv preprint arXiv:1911.09925}, vol.~3, no.~25, pp. 15--17, 2019.

\bibitem{wei2017automated}
X.~Wei, C.~H. Yu, P.~Zhang, Y.~Chen, Y.~Wang, H.~Hu, Y.~Liang, and J.~Cong,
  ``Automated systolic array architecture synthesis for high throughput cnn
  inference on fpgas,'' in \emph{Proceedings of the 54th Annual Design
  Automation Conference 2017}, 2017, pp. 1--6.

\bibitem{noll2023pangraph}
N.~Noll, M.~Molari, L.~P. Shaw, and R.~A. Neher, ``Pangraph: scalable bacterial
  pan-genome graph construction,'' \emph{Microbial Genomics}, vol.~9, no.~6, p.
  001034, 2023.

\bibitem{hickey2020genotyping}
G.~Hickey, D.~Heller, J.~Monlong, J.~A. Sibbesen, J.~Sir{\'e}n, J.~Eizenga,
  E.~T. Dawson, E.~Garrison, A.~M. Novak, and B.~Paten, ``Genotyping structural
  variants in pangenome graphs using the vg toolkit,'' \emph{Genome biology},
  vol.~21, pp. 1--17, 2020.

\bibitem{kung1979systolic}
H.~T. Kung and C.~E. Leiserson, ``Systolic arrays (for vlsi),'' in \emph{Sparse
  Matrix Proceedings 1978}, vol.~1.\hskip 1em plus 0.5em minus 0.4em\relax
  Society for industrial and applied mathematics Philadelphia, PA, USA, 1979,
  pp. 256--282.

\bibitem{brent1984systolic}
R.~P. Brent and H.-T. Kung, ``Systolic vlsi arrays for polynomial gcd
  computation,'' \emph{IEEE Transactions on Computers}, vol. 100, no.~8, pp.
  731--736, 1984.

\bibitem{gentleman1982matrix}
W.~M. Gentleman and H.~Kung, ``Matrix triangularization by systolic arrays,''
  in \emph{Real-time signal processing IV}, vol. 298.\hskip 1em plus 0.5em
  minus 0.4em\relax SPIE, 1982, pp. 19--26.

\bibitem{cong_fpga_2022}
\BIBentryALTinterwordspacing
J.~Cong, J.~Lau, G.~Liu, S.~Neuendorffer, P.~Pan, K.~Vissers, and Z.~Zhang,
  ``{FPGA} {HLS} {Today}: {Successes}, {Challenges}, and {Opportunities},''
  \emph{ACM Transactions on Reconfigurable Technology and Systems}, vol.~15,
  no.~4, pp. 51:1--51:42, Aug. 2022. [Online]. Available:
  \url{https://dl.acm.org/doi/10.1145/3530775}
\BIBentrySTDinterwordspacing

\bibitem{rupnowStudyHighlevelSynthesis2011}
K.~Rupnow, Y.~Liang, Y.~Li, and D.~Chen, ``A study of high-level synthesis:
  {{Promises}} and challenges,'' in \emph{2011 9th {{IEEE International
  Conference}} on {{ASIC}}}, Oct. 2011, pp. 1102--1105.

\bibitem{fahim_hls4ml_2021}
\BIBentryALTinterwordspacing
F.~Fahim, B.~Hawks, C.~Herwig, J.~Hirschauer, S.~Jindariani, N.~Tran, L.~P.
  Carloni, G.~Di~Guglielmo, P.~Harris, J.~Krupa, D.~Rankin, M.~B. Valentin,
  J.~Hester, Y.~Luo, J.~Mamish, S.~Orgrenci-Memik, T.~Aarrestad, H.~Javed,
  V.~Loncar, M.~Pierini, A.~A. Pol, S.~Summers, J.~Duarte, S.~Hauck, S.-C. Hsu,
  J.~Ngadiuba, M.~Liu, D.~Hoang, E.~Kreinar, and Z.~Wu,
  ``\BIBforeignlanguage{en}{hls4ml: {An} {Open}-{Source} {Codesign} {Workflow}
  to {Empower} {Scientific} {Low}-{Power} {Machine} {Learning} {Devices}},''
  Mar. 2021, arXiv:2103.05579 [physics]. [Online]. Available:
  \url{http://arxiv.org/abs/2103.05579}
\BIBentrySTDinterwordspacing

\bibitem{shahshahaniFrameworkModelingOptimizing2020}
M.~Shahshahani, B.~Khabbazan, M.~Sabri, and D.~Bhatia, ``A {{Framework}} for
  {{Modeling}}, {{Optimizing}}, and {{Implementing DNNs}} on {{FPGA Using
  HLS}},'' in \emph{2020 {{IEEE}} 14th {{Dallas Circuits}} and {{Systems
  Conference}} ({{DCAS}})}.\hskip 1em plus 0.5em minus 0.4em\relax Dallas, TX,
  USA: IEEE, Nov. 2020, pp. 1--6.

\bibitem{barenghiOpenCLHLSBased2018}
A.~Barenghi, M.~Madaschi, N.~Mainardi, and G.~Pelosi, ``{{OpenCL HLS Based
  Design}} of {{FPGA Accelerators}} for {{Cryptographic Primitives}},'' in
  \emph{2018 {{International Conference}} on {{High Performance Computing}} \&
  {{Simulation}} ({{HPCS}})}.\hskip 1em plus 0.5em minus 0.4em\relax Orleans:
  IEEE, Jul. 2018, pp. 634--641.

\bibitem{homsirikamolNewHLSbasedMethodology2017}
E.~Homsirikamol and K.~G. George, ``Toward a new {{HLS-based}} methodology for
  {{FPGA}} benchmarking of candidates in cryptographic competitions: {{The
  CAESAR}} contest case study,'' in \emph{2017 {{International Conference}} on
  {{Field Programmable Technology}} ({{ICFPT}})}.\hskip 1em plus 0.5em minus
  0.4em\relax Melbourne, VIC: IEEE, Dec. 2017, pp. 120--127.

\bibitem{hls_read_mapping}
D.~Castells-Rufas, S.~Marco-Sola, J.~C. Moure, Q.~Aguado, and A.~Espinosa,
  ``Fpga acceleration of pre-alignment filters for short read mapping with
  hls,'' \emph{IEEE Access}, vol.~10, pp. 22\,079--22\,100, 2022.

\bibitem{hls_chaining}
K.~Liyanage, H.~Gamaarachchi, R.~Ragel, and S.~Parameswaran, ``Cross layer
  design using hw/sw co-design and hls to accelerate chaining in genomic
  analysis,'' \emph{IEEE Transactions on Computer-Aided Design of Integrated
  Circuits and Systems}, vol.~42, no.~9, pp. 2924--2937, 2023.

\bibitem{hls_genome_assembly}
P.~Meng, M.~Jacobsen, M.~Kimura, V.~Dergachev, T.~Anantharaman, M.~Requa, and
  R.~Kastner, ``Hardware accelerated novel optical de novo assembly for
  large-scale genomes,'' in \emph{2014 24th International Conference on Field
  Programmable Logic and Applications (FPL)}, 2014, pp. 1--8.

\bibitem{benkrid_highly_2009}
\BIBentryALTinterwordspacing
K.~Benkrid, {Ying Liu}, and A.~Benkrid, ``\BIBforeignlanguage{en}{A {Highly}
  {Parameterized} and {Efficient} {FPGA}-{Based} {Skeleton} for {Pairwise}
  {Biological} {Sequence} {Alignment}},'' \emph{\BIBforeignlanguage{en}{IEEE
  Transactions on Very Large Scale Integration (VLSI) Systems}}, vol.~17,
  no.~4, pp. 561--570, Apr. 2009. [Online]. Available:
  \url{http://ieeexplore.ieee.org/document/4773142/}
\BIBentrySTDinterwordspacing

\end{thebibliography}

\end{document}